\documentclass[sigconf, screen]{acmart}  %
\usepackage{acmart-taps}

\AtBeginDocument{%
  }

\copyrightyear{2025}
\acmYear{2025}
\setcopyright{rightsretained}
\acmConference[UIST '25]{The 38th Annual ACM Symposium on User Interface Software and Technology}{September 28-October 1, 2025}{Busan, Republic of Korea}
\acmBooktitle{The 38th Annual ACM Symposium on User Interface Software and Technology (UIST '25), September 28-October 1, 2025, Busan, Republic of Korea}\acmDOI{10.1145/3746059.3747746}
\acmISBN{979-8-4007-2037-6/2025/09}

\usepackage{listings}            %
\usepackage{enumitem}            %

\definecolor{codegreen}{rgb}{0,0.6,0}
\definecolor{codegray}{rgb}{0.5,0.5,0.5}
\definecolor{codepurple}{rgb}{0.58,0,0.82}
\definecolor{backcolour}{rgb}{0.95,0.95,0.92}

\lstdefinestyle{mystyle}{
    backgroundcolor=\color{backcolour},   
    commentstyle=\color{codegreen},
    keywordstyle=\color{magenta},
    numberstyle=\tiny\color{codegray},
    stringstyle=\color{codepurple},
    basicstyle=\ttfamily\footnotesize,
    breakatwhitespace=false,         
    breaklines=true,                 
    captionpos=b,                    
    keepspaces=true,                 
    numbers=left,                    
    numbersep=5pt,                  
    showspaces=false,                
    showstringspaces=false,
    showtabs=false,                  
    tabsize=2
}

\newif\ifauthornotes
\newif\ifstrike
\newif\iftodo
\newif\ifrevise
\newif\ifadd
\newif\ifreplace

\newcommand{\strike}[1]{\ifstrike{\color{red}{\texorpdfstring{\sout{#1}}{#1}}}\fi}

\newcommand{\add}[1]{\ifadd{\leavevmode\color{magenta}{#1}}\else{#1}\fi}

\begin{document}

\title[{OnGoal}: Tracking and Visualizing Conversational Goals in Multi-Turn Dialogue\\ with Large Language Models]{{OnGoal}: Tracking and Visualizing Conversational Goals in Multi-Turn Dialogue with Large Language Models}

\author{Adam J Coscia}
\authornote{Work done during an internship at Adobe.}
\orcid{0000-0002-0429-9295}
\email{acoscia6@gatech.edu}
\affiliation{%
  \institution{Georgia Institute of Technology}
  \city{Atlanta}
  \state{Georgia}
  \country{USA}
}%

\author{Shunan Guo}
\authornote{Supervised work during internship.}
\orcid{0000-0001-5355-8399}
\email{sguo@adobe.com}
\author{Eunyee Koh}
\orcid{0000-0003-2091-5972}
\email{eunyee@adobe.com}
\affiliation{%
  \institution{Adobe Research}
  \city{San Jose}
  \state{California}
  \country{USA}
}%

\author{Alex Endert}
\orcid{0000-0002-6914-610X}
\email{endert@gatech.edu}
\affiliation{%
  \institution{Georgia Institute of Technology}
  \city{Atlanta}
  \state{Georgia}
  \country{USA}
}%

\renewcommand{\shortauthors}{Coscia et al.}

\begin{abstract}
  As multi-turn dialogues with large language models (LLMs) grow longer and more complex, how can users better evaluate and review progress on their conversational goals?
  We present OnGoal, an LLM chat interface that helps users better manage goal progress.
  OnGoal provides real-time feedback on goal alignment through LLM-assisted evaluation, explanations for evaluation results with examples, and overviews of goal progression over time, enabling users to navigate complex dialogues more effectively.
  Through a study with 20 participants on a writing task, we evaluate OnGoal against a baseline chat interface without goal tracking.
  Using OnGoal, participants spent less time and effort to achieve their goals while exploring new prompting strategies to overcome miscommunication, suggesting tracking and visualizing goals can enhance engagement and resilience in LLM dialogues.
  Our findings inspired design implications for future LLM chat interfaces that improve goal communication, reduce cognitive load, enhance interactivity, and enable feedback to improve LLM performance.
\end{abstract}



\begin{teaserfigure}
  \includegraphics[width=\textwidth]{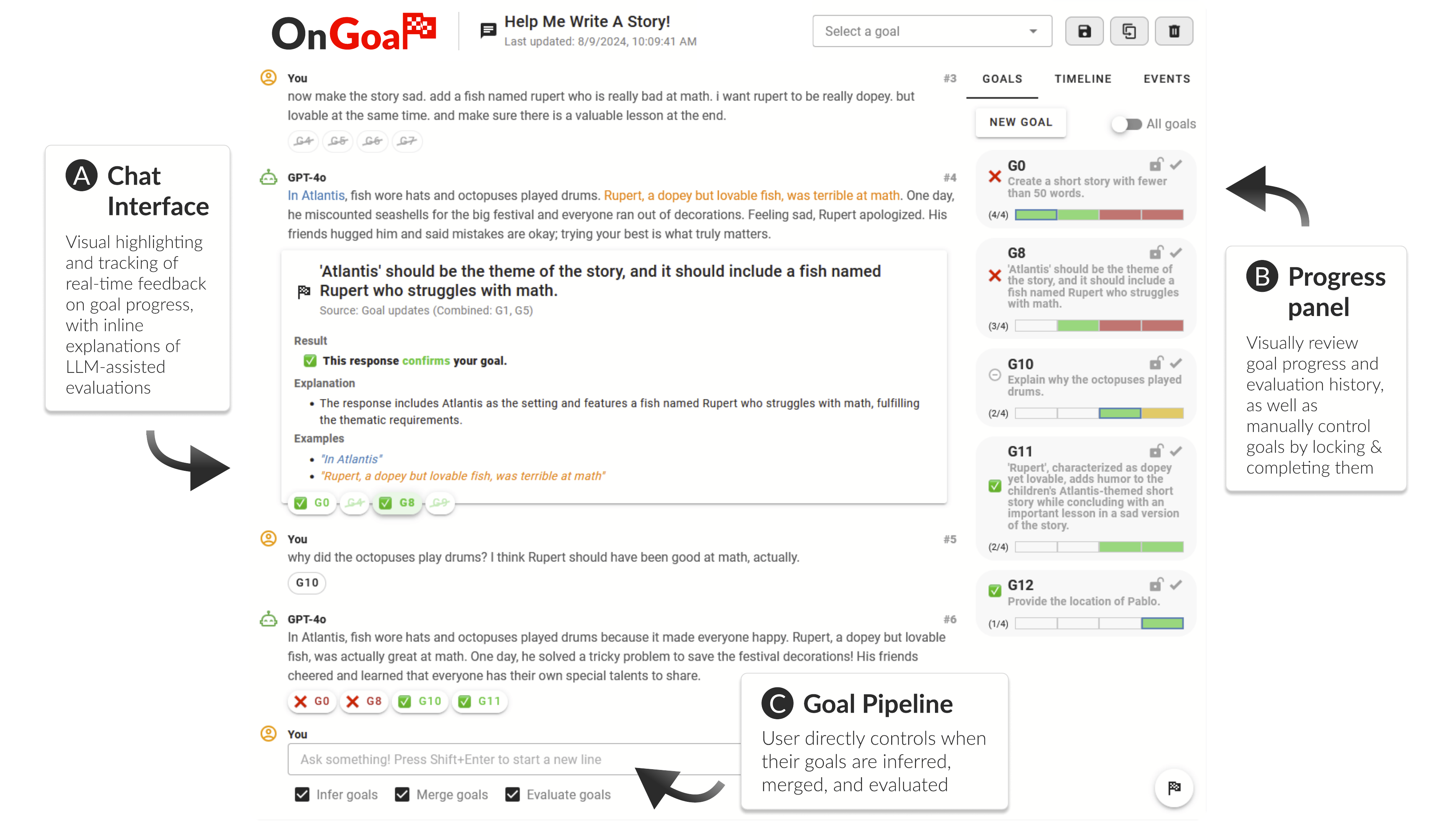}
  \caption{%
    OnGoal tracks and visualizes conversational goals such as requests and suggestions in multi-turn dialogue with LLMs, helping users better evaluate and review their goal progress.
  }%
  \Description{%
    Web page interface of VisPile.
    The page consists of a small header with title, a left main content area with a scrolling chat interface and text input at the bottom, and a right side bar with three tabs: Goals, Timeline, and Events.
  }%
  \label{fig:teaser}
\end{teaserfigure}

\maketitle

\section{Introduction}
\label{sec:introduction}

Large language models (LLMs) have significantly improved their ability to handle multi-turn, text-based interactions that span longer, more complex conversations \cite{Zheng:2023:ChatbotArena, duan2023botchat}.
This has led users to engage actively in dyadic (two-agent) conversational turn-taking with LLMs similar to human-human conversations, using conversational goals such as questions, requests, offers, or suggestions to structure communication \cite{Stivers:2009:TurnTaking}.
While multi-turn interactions have been shown to increase expressivity and flexibility \cite{Ross:2023:ProgrammersAssistant}, the length and shifting context of back-and-forth conversations can make it challenging for users to evaluate and review their conversational goals as they evolve over time.
For example, users may struggle with under-specified or conflicting goals, parsing long chats for progress, or addressing stagnant and forgotten goals \cite{Kim:2024:DissatisfactionGPTTypes}.
These challenges often cause users to repeat prompts, ignore their goals, or restart the conversation entirely, losing valuable progress and insights gained along the way \cite{Liang:2023:UsabilityAIAssistants}.

Managing goals that evolve over time or span multiple turns is challenging, particularly with linear chat interfaces.
At the same time, these interfaces remain widely adopted for interacting with LLMs, offering a familiar and intuitive structure for multi-turn conversations.
For instance, an analyst using Copilot to explore a sales dataset might start with a broad goal --- like identifying overall sales trends --- and progressively refine their goal to investigate seasonal patterns or identify outliers.
Yet over several turns, it can be difficult for the analyst to \textit{evaluate} if the LLM's responses address their current goal, or if the model is still addressing prior goals without clear transitions.
Alternatively, a wedding planner coordinating activities and themes with ChatGPT may be more concerned with \textit{reviewing} how goals have been addressed over time.
As requests and suggestions accumulate, it can become difficult to trace which goals have been fulfilled, which remain pending, or whether new requests contradict with earlier ones as they scroll an ever-growing chat log.
By refining the linear chat paradigm, we aim to balance familiarity with functionality, making AI-assisted conversations more effective and user-friendly.
These challenges inspire us to ask:

\begin{quote}
    \textbf{``How can a linear chat interface help users evaluate and review their conversational goals over extended dialogue with LLMs?''}
\end{quote}

To address \add{this challenge}\strike{ our question}, we \add{developed}\strike{ introduce} OnGoal, a chat interface augmented with goal-tracking visualizations to assist users in evaluating and refining their conversational objectives.
Our approach is grounded in insights from ML, HCI and visualization literature, which we synthesized into three design challenges around conversing with LLMs.
In response to these challenges, OnGoal integrates a LLM-assisted \textbf{goal pipeline} to infer, merge, and evaluate a user's goals against LLM responses over time.
The \textbf{chat interface} visualizes goal progress inline with explanations of how an LLM response addresses each goal, as well as on the side in multiple progress summary views, increasing user awareness on goal alignment throughout long, complex LLM responses.
Across these views, OnGoal provides \textbf{text highlighting} to help users compare messages over time for key insights related to LLM behaviors including distractions, context switching, and topic drift.

To assess OnGoal's effectiveness, we conducted a user study with 20 participants, \add{focusing on LLM-assisted writing tasks as a representative scenario for multi-turn, goal-oriented interactions. We compared}\strike{ comparing} how well \add{OnGoal}\strike{ it} supports goal evaluation and \add{review compared to}\strike{ reviewing in multi-turn, LLM-assisted writing tasks versus} a baseline chat interface without goal tracking or visualizations.
\add{We examined the effectiveness of OnGoal's features in supporting task outcomes and user experience. Our study also surfaced behavioral insights into how users adapt their communication strategies when interacting with goal-feedback visualizations.}
Overall, OnGoal encouraged new strategies for communicating goals with the LLM, shifting the time and effort spent away from reading the chat and towards evaluating and reviewing goals, and a greater understanding of different LLM behaviors that could cause conversations to derail.
Our insights led us to synthesize design implications for future interfaces that help users better evaluate and review conversational goals: (1) provide multiple methods for goal communication, such as setting goals up front or letting them be inferred automatically; (2) enhance interactivity with visual and directive tools, such as letting the user indicate where they want the LLM to focus attention or explain itself; (3) reduce cognitive load with proactive goal tracking by including goal alerts and progress snapshots; and (4) enable feedback on how goals are addressed to improve LLM evaluations and examples.

In summary, we contribute:

\strike{Design challenges for addressing challenges in managing conversational goals during multi-turn dialogue with LLMs.}

\begin{itemize}[topsep=1pt]
    \item OnGoal, a chat interface augmented with in-situ visualizations for tracking conversational goals\add{, developed to address design design challenges we identified in managing conversational goals during multi-turn dialogues with LLMs}.
    \item A user study with 20 participants \add{evaluating OnGoal's effectiveness in helping}\strike{ showing how OnGoal helps} users manage conversational goals in LLM conversations compared to a baseline chat interface\add{, and revealing how real-time goal feedback shapes users' communication strategies}.
    \item Design implications for future LLM chat interfaces that improve evaluating and reviewing conversational goals.
\end{itemize}

\section{Related Work}
\label{sec:related_work}

This work leverages LLMs as conversational agents, defined as AI systems that generate multi-turn, open-domain dialogue dynamically based on a user's contextual input.
Existing non-LLM agents like Siri and Alexa engage in unstructured, back-and-forth dialogue, supporting long multi-turn conversations to accomplish tasks like answering questions \cite{Deriu:2021:SurveyDialogueEvaluation}.
While agents can be mixed-initiative, we focus on augmenting user-initiated interactions.
In this section, we discuss related work on interacting with LLMs as conversational agents, sensemaking of LLM conversations, and how visualizations help users better understand LLMs during multi-turn dialogue.

\subsection{LLM-Based Conversational Agents}
\label{sec:llm_conversation}
Increased context lengths have improved LLMs' ability to understand and respond to multiple user prompts in sequence, enabling new use cases for conversational interaction \cite{Zheng:2023:ChatbotArena}.
In tasks like writing and information foraging, conversations often remain $10$ turns or less due to mental fatigue from an ever-growing chat log \cite{Huang:2024:ConvLengthLLMStudy}.
Gao et al. reviewed HCI literature on human-LLM interactions, synthesizing four phases in which LLM assistance occurs; (1) planning the goals of the conversation; (2) facilitating a new interaction; (3) refining an existing interaction; (4) testing interactions \cite{Gao:2024:HumanLLMInteractionModes}.
When people engage with LLMs in multi-turn conversation for help tasks like programming assistance, Ross et al. discovered that multi-turn conversations can enhance co-creativity, boost engagement and productivity, and increase resilience to errors over single-turn interactions \cite{Ross:2023:ProgrammersAssistant}.
Based on how LLMs are used as conversational agents, we aim to improve the linear chat interface experience when users converse with an LLM in a multi-turn setting to plan, execute, and refine their goals towards accomplishing a specific task.

Despite the growing use of LLMs in multi-turn conversation, LLMs have limitations in handling long conversations.
Kim et al. studied dissatisfaction types of GPT responses and found challenges around intent understanding, depth and originality, accuracy, transparency, refusal to answer, ethics and integrity, format and attitude \cite{Kim:2024:DissatisfactionGPTTypes}.
LLMs can exhibit undesirable behaviors that can derail the conversation, leading to distractions \cite{Shi:2023:LLMsDistracted}, task switching \cite{Gupta:2024:LLMTaskSwitching} or topic drift \cite{Ashby:2024:LLMConversationTopic, Liu:2024:LostInTheMiddle}.
A major concern is the tendency of LLMs to forget context from earlier turns, particularly in long conversations.
For instance, Liu et al. discovered that LLMs such as GPT-3.5 tend to omit information in the middle of requests, as context windows increase in length \cite{Liu:2024:LostInTheMiddle}.
These limitations make it challenging for users to maintain coherent and effective conversations over time.

Our work seeks to address these limitations in the context of conversational goal management.
In natural language processing (NLP), dialogue state tracking (DST) has been widely used to model user goals and track conversation states, helping analysts uncover valuable interaction patterns \cite{kulkarni2024synthdst, niu2024enhancing}.
DST is typically employed in post-conversation analysis as a classification task, where the space of goals or states is predefined and models are trained on labeled data.
However, DST is less suited for dynamic, context-driven conversations with LLMs, where tasks are dynamic and continuously evolving.
Because DST is primarily retrospective, it also fails to address users' real-time challenges such as maintaining context, understanding response relevance, and tracking shifts in conversation flow.
To bridge this gap, our work aims to enhance user awareness of conversation goals and states during interactions, providing real-time support to improve communication and sensemaking.

\subsection{Sensemaking of LLM Conversations}
\label{sec:llm_sensemaking}
Understanding long, text-based conversations is inherently challenging.
Conversational agents can fail to address all parts of a user's request, lose context or drift as the conversation progresses, or provide contradictory responses when the user seeks clarification \cite{Higashinaka:2021:TaxonomyErrorsDialogue}.
\add{Prior work in task decomposition and chain-of-thought prompting sought to improve model reasoning and verification by breaking down tasks into sub-tasks or reasoning steps~\cite{khot2022decomposed, wei2022chain}. However, such approaches are predominantly model-centric, aiming to improve a model's internal reasoning and output verification. Task decomposition and evaluation typically occur within the model itself, with intermediate reasoning steps or sub-goals not explicitly surfaced to end users. In contrast, our work focuses on supporting users in monitoring and understanding goal progress as they interact, addressing the complementary challenge of user-facing feedback in multi-turn conversations.}

\add{Addressing the lack of user-facing support contributes to broader challenges in}\strike{ Such issues complicate} how users engage with and trust LLMs when accomplishing tasks \cite{Papenmeier:2022:TrustAIExplanations}.
For example, Liang et al. conducted a survey on the usability of LLMs as conversational assistants \cite{Liang:2023:UsabilityAIAssistants}.
They found that LLM responses can be excessively long and contain too many terms and structures, making them time-consuming to read and difficult to track over time in multi-turn settings, as well as making it unclear how responses connect between prompts.
These issues can hinder users' ability to understand if their goals were satisfied during conversation with LLMs.

To improve the usability of LLM-based conversation agents, researchers have explored methods to provide real-time feedback.
Explanations of AI responses have been widely used as a critical method to calibrate trust in AI models like LLMs \cite{Papenmeier:2022:TrustAIExplanations}.
For example, OpenAI released CriticGPT, a model to provide critic feedback on ChatGPT responses during a user's conversational session \cite{McAleese:2024:CriticGPT}.
Hernandez-Bocanegra and Ziegler found that interactive, graphical explanations of LLM behaviors can lead to more positive user experiences \cite{Hernandez-Bocanegra:2023:ExplainingRecsInConvo}.
Gero et al. studied how visual highlighting techniques could be used to support sensemaking of LLM text at scale \cite{Gero:2024:SensemakingLLMsAtScale}.
Inspired by these findings, this work enhances LLM-human communication by visualizing goal feedback across messages over time as a way to help users make sense of their progress.

\subsection{Visualizing LLM Conversations}
\label{sec:llm_visualization}
The visualization community has a rich history of developing methods to support the analysis and sensemaking of text conversations \cite{Tat:2002:VisualizingHumanDialogue, Venolia:2003:EmailConversationVis, Baker:2009:VisEnhancesSensemaking}.
Many prior works focus on human-to-human conversations and post-hoc analysis of interaction patterns \cite{Fu:2018:TCal, Wang:2021:DiscussionFlows}.
For example, StuGPTViz visualizes conversations between ChatGPT and students after the student has already conversed with the LLM \cite{Chen:2025:StuGPTViz}.
Few works have explored visualizing interaction patterns from an LLM conversation in real time, and fewer still have explored this idea to help users manage their conversational goals.

To enhance LLM conversations, many tools like PromptChainer \cite{Tongshuang:2022:PromptChainer} and PromptAid \cite{Mishra:2023:PromptAid} focus on refining prompts to optimize single-turn responses.
Alternatively, visualizations can also improve multi-turn conversations by helping users better manage their goals over time.
For example, Hong et al. developed AI Threads as a multi-threaded approach to managing conversational context and improving LLM responses \cite{Hong:2023:ConversationalAIThreads}.
Suchmann et al. developed a GUI for visualizing the branching topics of conversation with a recommendation system over time, helping users review progress \cite{Suchmann:2023:BranchingConvoInSituVis}.
Going further, a few systems like Graphologue \cite{Jiang:2023:Graphologue} and Sensecape \cite{Suh:2023:Sensecape} aid sensemaking of multi-turn conversations with LLMs using interface designs that break the linear chat format.
These tools offer structured ways to manage dialogue flow and let the user perform follow-up interactions through direct manipulation, helping explain or explore LLM responses.

\add{These systems introduce new ways to navigate responses and enhance exploration. To do this, they focus on structuring outputs, rather than surfacing the user's own conversational goals. In contrast, OnGoal introduces a user-facing layer of inferred conversational goals, towards helping users monitor evolving objectives they pose to the LLM.}
For example, the non-linear chat structures \add{introduced in Graphologue and Sensecape} offer powerful sensemaking capabilities, yet they also introduce added complexity that may disrupt users' mental models and require a steep learning curve.
\add{OnGoal instead extends the familiar linear chat interface, embedding goal tracking and visualization directly within the conversational flow to support goal management without disrupting established interaction patterns. Our work contributes to the space of}\strike{ Few recent works have explored design opportunities for} augmenting multi-turn, conversational interactions in linear chat interfaces \cite{Bursztyn:2021:LLMConvoRec, Stigall:2023:ChatbotsVisualEDA}\add{, focusing on supporting users in managing conversational goals throughout extended dialogues}\strike{ We aim to enhance linear interfaces with improved goal visualization and tracking features, preserving its intuitive flow while addressing key limitations}.

\section{Designing Goal Tracking and Visualization}
\label{sec:methodology}

To formalize our design rationale, we organized insights from related ML, HCI and visualization literature on conversing with LLMs.
The advent of LLMs that can handle longer inputs has inspired the HCI and visualization communities to develop and study chat interfaces with LLMs as dyadic conversational agents \cite{Kim:2021:ReviewDyadicConvVis}.
However, several ML studies reveal issues that LLMs have in handling multi-turn dialogue, including context switching \cite{Gupta:2024:LLMTaskSwitching} and topic drift \cite{Ashby:2024:LLMConversationTopic, Liu:2024:LostInTheMiddle}.
HCI literature further reveals usability challenges of LLMs as conversational agents \cite{Liang:2023:UsabilityAIAssistants} as well as dissatisfaction in how LLMs address users during conversations \cite{Kim:2024:DissatisfactionGPTTypes}.
Inspired by this growth and the challenges it poses, we distilled the issues that users might face when interacting with LLMs in multi-turn conversation into key design challenges \textbf{(C)} for our system to address.
The challenges are integrated into our descriptions of the system (Sect.~\ref{sec:system}) and evaluation (Sect.~\ref{sec:evaluation}) throughout the rest of the paper.

\medskip
\noindent\textbf{C1.  }
\textit{With many overlapping conversational goals, LLMs can miss what the user wants.}
One common reason for dissatisfaction with LLMs as conversational agents is their difficulty in accurately discerning user intent.
This often arises when responses focus on unexpected parts of a user's message, overlook key user goals, struggle to address conflicting goals, or misinterpret what the user meant \cite{Kim:2024:DissatisfactionGPTTypes}.
For example, helping a user write a story may require the LLM to balance explicit requests, such as using both formal and informal language, replacing out-dated goals like including more imagery as the story becomes overly saturated, and merging conflicting suggestions.
However, a growing conversation can cause the LLM to forget or ignore previous user requests \cite{Liu:2024:LostInTheMiddle}, further exacerbating problems if prompting continues.
Without clear explanations, users may struggle with balancing prompt length and clarity, leading to repeatedly prompting the same message and ultimately getting frustrated and losing confidence when the LLM fails to adapt.
A chat interface should formalize the communication of the user's goals and explain the LLM's understanding and response to user messages, highlighting under-specified, misinterpreted, or ignored goals.
This would help users refine their input and reduce communication breakdowns.

\medskip
\noindent\textbf{C2.  }
\textit{As messages accumulate, sensemaking across long, complex LLM responses becomes effortful and time-consuming.}
Even when the LLM correctly interprets user goals, tracking the progress within an increasingly long chat log can become untenable.
LLM responses are often overly long and filled with irrelevant ``fluff'', introducing extraneous terms and themes that make it hard to track goal progress \cite{Liang:2023:UsabilityAIAssistants}.
This often compounds with the general challenges of dialogue systems repeating themselves, losing context over time, or contradicting themselves during clarifications \cite{Higashinaka:2021:TaxonomyErrorsDialogue}.
Challenges making sense of the text begin to snowball: evaluating multiple goals against each response, reviewing the progress of goals across responses and determining when goals are fully addressed require more time and effort.
These obstacles can shift users' attention away from their task and ultimately reduce their confidence \cite{Lee:2025:GenAICriticalThinking}.
Summarizing goal status and tracking progress over time can alleviate this burden, reducing the need for users to manually sift through the entire chat log. Additionally, such summaries could surface patterns in LLM behavior, helping users to adjust their prompting strategies more effectively. 

\aptLtoX[graphic=no, type=html]{
}{
\begin{figure*}[!t]
  \centering
  \includegraphics[width=\linewidth]{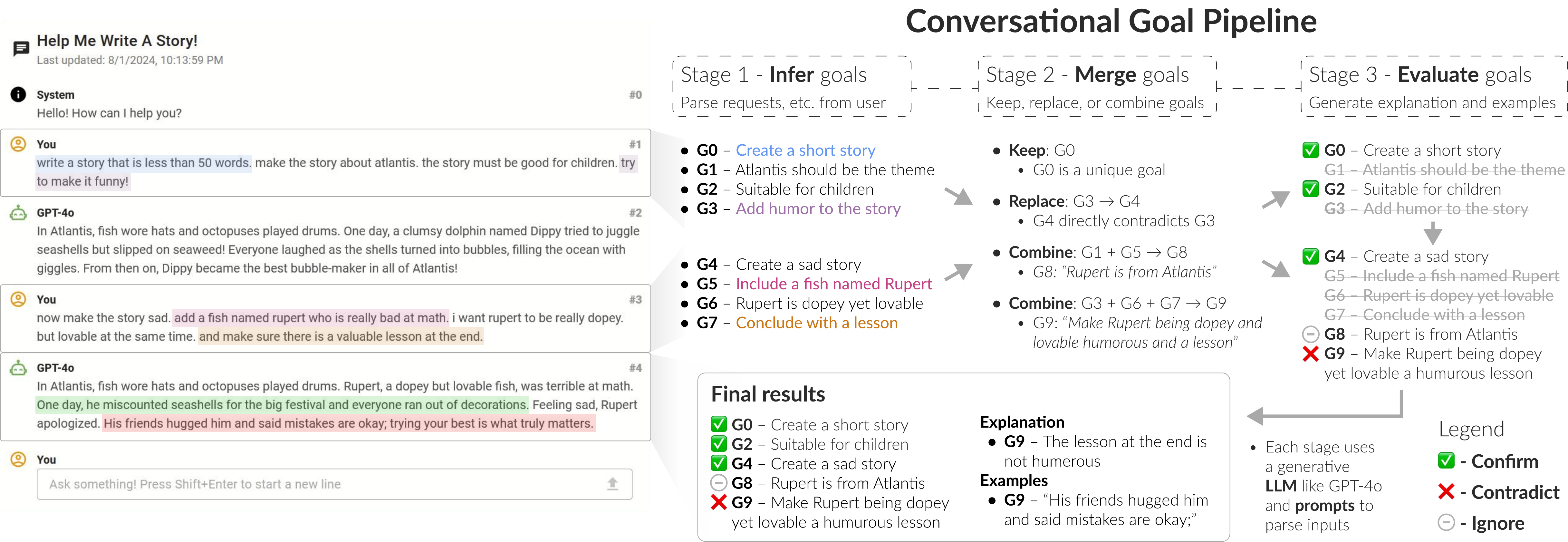}
  \caption{%
    An example of the conversational goal pipeline in action.
    OnGoal uses a generative LLM (e.g., GPT-4o) and prompt engineering to infer, merge, and evaluate goals.
    The results are then visualized in the chat.
  }%
  \Description{%
    A chat interface screenshot on the left, and a framework diagram on the right with three stages.
    In the framework, from left to right: Stage 1, Infer goals; Stage 2, Merge goals; Stage 3, Evaluate goals.
    Below the stages on the right is a Final Results stage.
    Every framework stage lists goals from the chat interface screenshot on the left as they are processed in each stage sequentially.
  }%
  \label{fig:goal_pipeline}
\end{figure*}
}

\medskip
\noindent\textbf{C3.  }
\textit{When the conversation derails, undesirable LLM behaviors are often opaque.}
Sometimes LLMs can fail to address goals in ways that are not immediately clear to the user.
For example, LLMs may forget conversational goals over time, get stuck on a specific goal, or address the goal in unexpected ways, like generating multiple versions of text or only applying the goal to a subsection of the response \cite{Deriu:2021:SurveyDialogueEvaluation}.
These undesirable behaviors can derail the conversation, leading to distractions \cite{Shi:2023:LLMsDistracted}, task switching \cite{Gupta:2024:LLMTaskSwitching} or topic drift \cite{Ashby:2024:LLMConversationTopic, Liu:2024:LostInTheMiddle}.
While model explanations can increase users' trust \cite{Papenmeier:2022:TrustAIExplanations}, identifying inconsistent behaviors from explanations is tricky with just text alone.
As a conversation grows, it becomes harder to track how the LLM addressed goals across multiple messages.
Visual strategies, such as highlighting relevant text to reveal goal alignment, or comparing key phrases across messages, could help users detect misalignment and help them regain control over the conversation \cite{Gero:2024:SensemakingLLMsAtScale}. 
Additionally, extracting examples within a message that exemplify where goal progress has been made and identifying recurring patterns across messages can further support users in understanding how the LLM is or is not addressing their goals.

\section{The {OnGoal} System}
\label{sec:system}

Based on our design challenges (Sect.~\ref{sec:methodology}), we developed \textit{OnGoal}, a chat interface augmented with in-situ visualizations for real-time, human-in-the-loop control over setting and tracking conversational goals.
Our system integrates a goal pipeline with multiple chat views and text highlighting techniques that help users evaluate and review conversational goals as they converse with an LLM in multi-turn dialogue.

The \textit{goal pipeline} (Sect.~\ref{sec:goal_pipeline}) infers, merges, and evaluates goals against LLM responses, and provides explanations and examples for how each goal was evaluated to increase communication and transparency.
The pipeline is embedded in a chat user interface featuring several visualizations (Sect.~\ref{sec:interface_views}) designed to summarize and compare goal progress, including inline goal glyphs, ex-situ timeline views of goal progress, and detailed views of individual goal progress.
Finally, OnGoal provides several text highlighting features (Sect.~\ref{sec:text_highlighting}) that show examples of goal alignment inline in the text, as well as comparison views of key phrases, similar and unique sentences between responses, helping users identify undesirable LLM behaviors to address.

\add{Each of OnGoal's views addresses a unique design challenge in general, domain-agnostic goal-tracking that the other views do not. Inline goal glyphs visualize how the LLM interpreted the user with explanations (\textbf{C1}), so users know what the LLM may be focusing on. Other views lack local explanations for LLM behaviors. As goals evolve over time, the ex-situ progress panel summarizes temporal trends in goal tracking (\textbf{C2}) that users would otherwise have to scroll up and down to find --- another unique feature. Finally, text highlighting adds concrete examples of LLM behaviors across messages (\textbf{C3}), where other views only address goal completion in each message individually. By combining their unique capabilities, the views in tandem aim to mitigate issues of cognitive overload that users often face when managing longer, multi-turn LLM dialogues.}

\subsection{Modeling Goals Using a Pipeline}
\label{sec:goal_pipeline}

\aptLtoX[graphic=no, type=html]{
\begin{figure*}[!t]
  \centering
  \includegraphics[width=\linewidth]{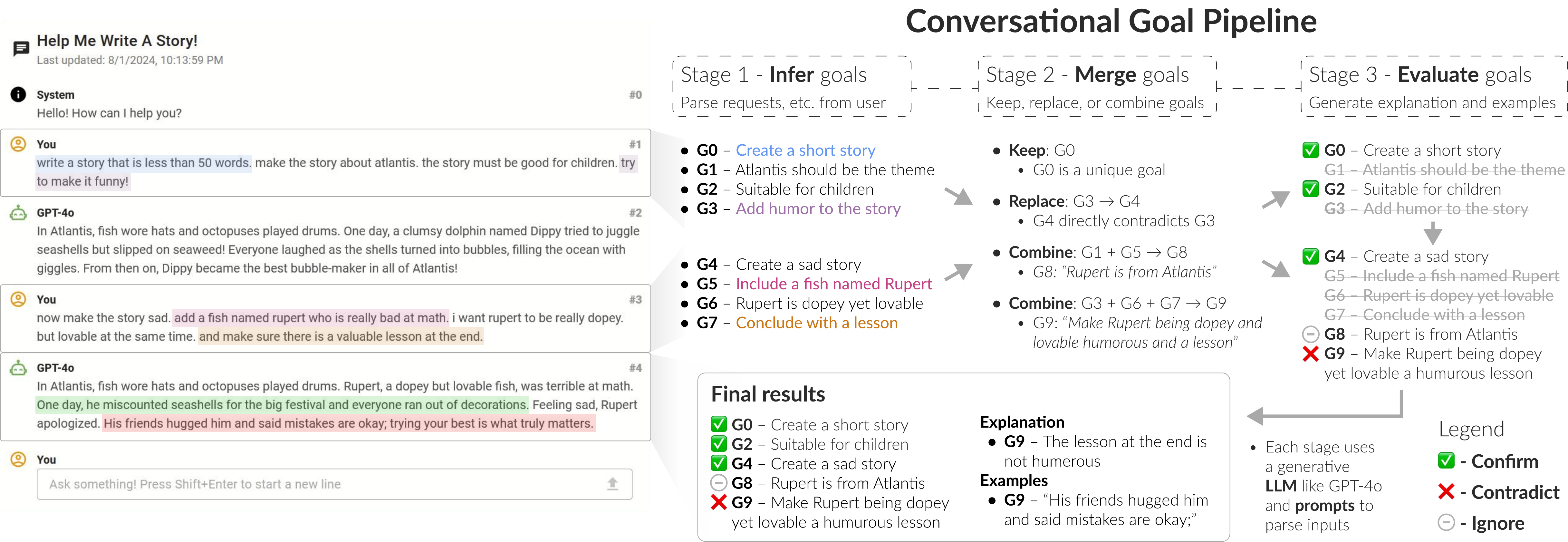}
  \caption{%
    An example of the conversational goal pipeline in action.
    OnGoal uses a generative LLM (e.g., GPT-4o) and prompt engineering to infer, merge, and evaluate goals.
    The results are then visualized in the chat.
  }%
  \Description{%
    A chat interface screenshot on the left, and a framework diagram on the right with three stages.
    In the framework, from left to right: Stage 1, Infer goals; Stage 2, Merge goals; Stage 3, Evaluate goals.
    Below the stages on the right is a Final Results stage.
    Every framework stage lists goals from the chat interface screenshot on the left as they are processed in each stage sequentially.
  }%
  \label{fig:goal_pipeline}
\end{figure*}
}{
}

\textit{OnGoal} utilizes a three-stage \textit{goal pipeline} to model conversational goals (Fig.~\ref{fig:goal_pipeline}).
Each stage of the pipeline is run by prompting a generative LLM (OpenAI's GPT-4o in our implementation) with goals from prior stages.
The pipeline LLM is independent of the chat LLM that the user is conversing with in the interface.
By running a goal pipeline, OnGoal can infer goals, merge similar goals, and evaluate goals against each LLM response.
The evaluations also generate explanations for why the LLM gave the evaluation it did, towards clarifying what the LLM understood about the user's goals and how it addressed them (\textbf{C1}).

\add{To create our prompts,} we defined conversational goals as \textbf{questions, requests, offers, or suggestions} from the user that an LLM should respond to in turn-taking order \cite{Stivers:2009:TurnTaking}.
\add{We include this definition in our prompts verbatim. The prompt also assigns one of the four types to each goal inferred.}
For example, a user might prompt the LLM, \textit{``I want to write a story''}, which the goal pipeline would infer as a request. The user could suggest, \textit{``You should make the story happy''}, offer a critique, \textit{``I think the story should be longer''}, or ask a question, \textit{``Why did you include a dog in the story?''} All these example clauses are conversational goals.
\add{Our definition is tailored to LLM-assisted writing as the primary use case in this work. Further, limiting goal inference to these four goal types helped the pipeline successfully infer, merge, and evaluate goals in the context of writing tasks, such as requesting feedback, suggesting edits, and asking questions. However, the definition of conversational goals can vary across real-world applications, which may require adaptation of goal types and inference prompts to fit different domains.}\strike{ OnGoal can extract multiple conversational goals, sometimes conflicting, within a single user message. The prompts for each stage of the pipeline all utilize this definition of conversational goals.}
Our prompts are included in the \add{appendix (Sect.~\ref{sec:llm_prompts})}\strike{ supplemental material}.

\aptLtoX[graphic=no, type=html]{
}{
\begin{figure*}[!t]
  \centering
  \includegraphics[width=\linewidth]{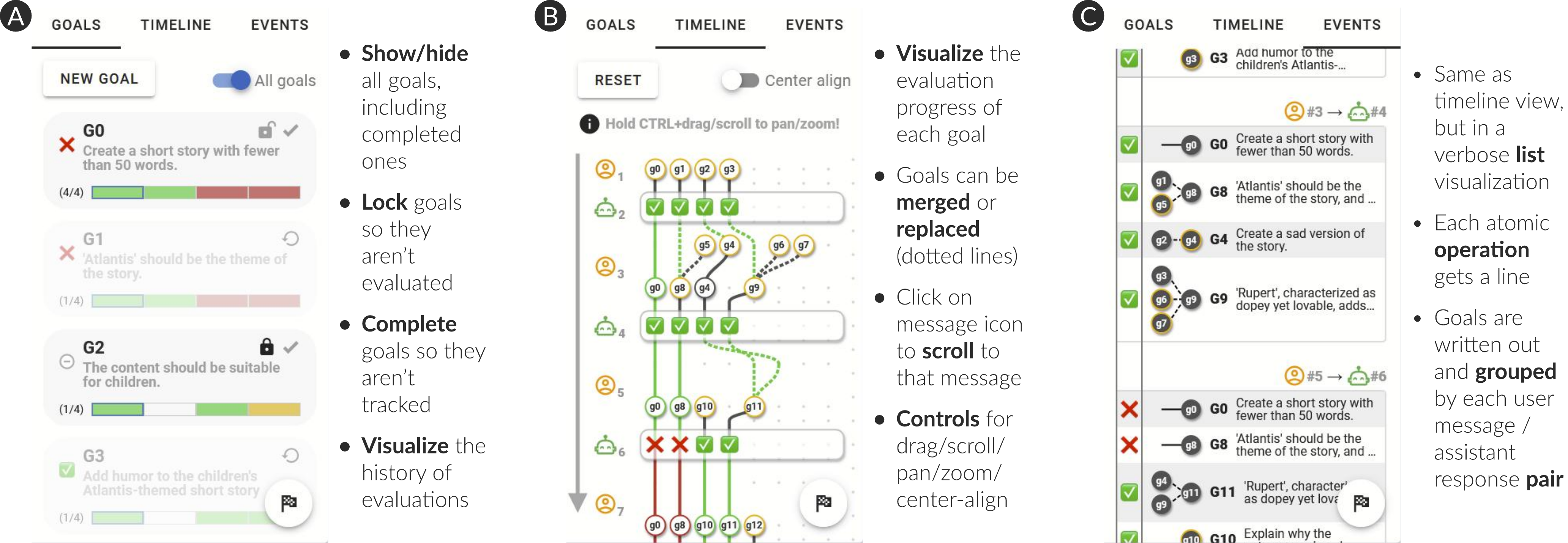}
  \caption{%
    The three tabs in the progress panel.
    In the goals tab (\textbf{A}), users can control goals by locking or completing them, create their own goals, and visualize a goal's evaluation history.
    In the timeline tab (\textbf{B}), the goal pipeline is visualized as a node-link diagram, where nodes flow top to bottom and are colored by their evaluation.
    In the events tab (\textbf{C}), the goal pipeline operations are visualized in a list view with more verbose descriptions.
    Users can brush and link a row in the timeline or events view with its corresponding message in the chat by clicking the numbered user icons.
  }%
  \Description{%
    Three sub-figures labeled A, B, and C.
    Each sub-figure shows a screenshot of the side bar on the right of the OnGoal web interface.
    A shows Goals tab, B shows the Timeline tab, and C shows the Events tab.
  }%
  \label{fig:progress_view}
\end{figure*}
}

The pipeline performs the following operations in order (see Fig.~\ref{fig:goal_pipeline} for an example).
Each step can be independently toggled on or off from the OnGoal interface (Fig.~\ref{fig:teaser}C).

\begin{enumerate}
    \itemsep0.5em
    \item \textbf{Infer. } The pipeline takes the user's message as input and infers conversational goals, which include \textit{questions, requests, offers, or suggestions}. These goals are stored as a list and passed to subsequent steps of analysis.
    \item \textbf{Merge. } In this step, the pipeline analyzes the existing list of goals from conversation history and compares them with the newly inferred goals from the latest interaction. It then performs one of the three operations: \textit{combine} similar existing and inferred goals, \textit{replace} an existing goal with a similar or contradictory inferred goal, or \textit{keep} unique existing and inferred goals.
    \item \textbf{Evaluate. } The pipeline takes the final list of merged goals, as well as the chat LLM's response to the user's message as input. It then evaluates each goal against the chat LLM's response. This prompt is used to determine if the response \textit{confirms}, \textit{contradicts}, or \textit{ignores} the goal, provide an explanation for the evaluation, and extract phrases from the LLM response as supporting evidence.
\end{enumerate}

\add{Our pipeline focuses on global goal tracking and support, i.e., parsing and applying goals to an entire LLM response. In the context of writing tasks, global goal tracking is useful to address specific issues in writing. For example, conflicting writing styles can cause the LLM to prioritize goals inconsistently. A global perspective can help users understand opaque LLM behaviors (\textbf{C3}), such as the LLM ignoring certain goals due to repeated language patterns that only appear over time. In dialogue state tracking, other types of goal support exist; e.g., fine-grained or local goal support, such as a user asking the LLM to edit the tone of a particular paragraph or sentence differently from the global tone. We discuss future work exploring local goal support in Sect.~\ref{sec:limitations_future}.}

\subsection{Visualizing Goals in the Chat UI}
\label{sec:interface_views}
OnGoal visualizes goal data from each stage of the pipeline in the chat interface using both in-situ evaluations and ex-situ summary views.
By integrating insights and patterns directly into the interface, OnGoal aims to improve communication (\textbf{C1}) while reducing the time and effort it takes to make sense of the chat history (\textbf{C2}).

\subsubsection{In-situ goal evaluations}
\label{sec:insitu_evaluations}
OnGoal presents two inline visuals within the chat messages themselves (Fig.~\ref{fig:teaser}A).
First, \textbf{goal glyphs} are shown under each message in the chat window to summarize inferred and evaluated goals.
Below a user's prompt, they indicate the goals that were inferred, while below an LLM's response, they reflect the final merged goals and their evaluation results colored accordingly -- green for confirm, red for contradict, and yellow for ignore.
This helps users quickly assess how goals were evaluated at a glance and evaluate how the LLM responses addressed their conversational goals over time as they scroll the conversation (\textbf{C2}).
Second, clicking on a goal glyph opens a more-detailed \textbf{goal explanation} inline.
For user goal glyphs, the view explains how the goal was inferred and highlights the users' phrase that the goal is derived from.
For LLM goal glyphs, the panel includes explanations of why the LLM assigned that evaluation as well as supporting evidence extracted from the LLM response. (\textbf{C1}).

\aptLtoX[graphic=no, type=html]{
\begin{figure*}[!t]
  \centering
  \includegraphics[width=\linewidth]{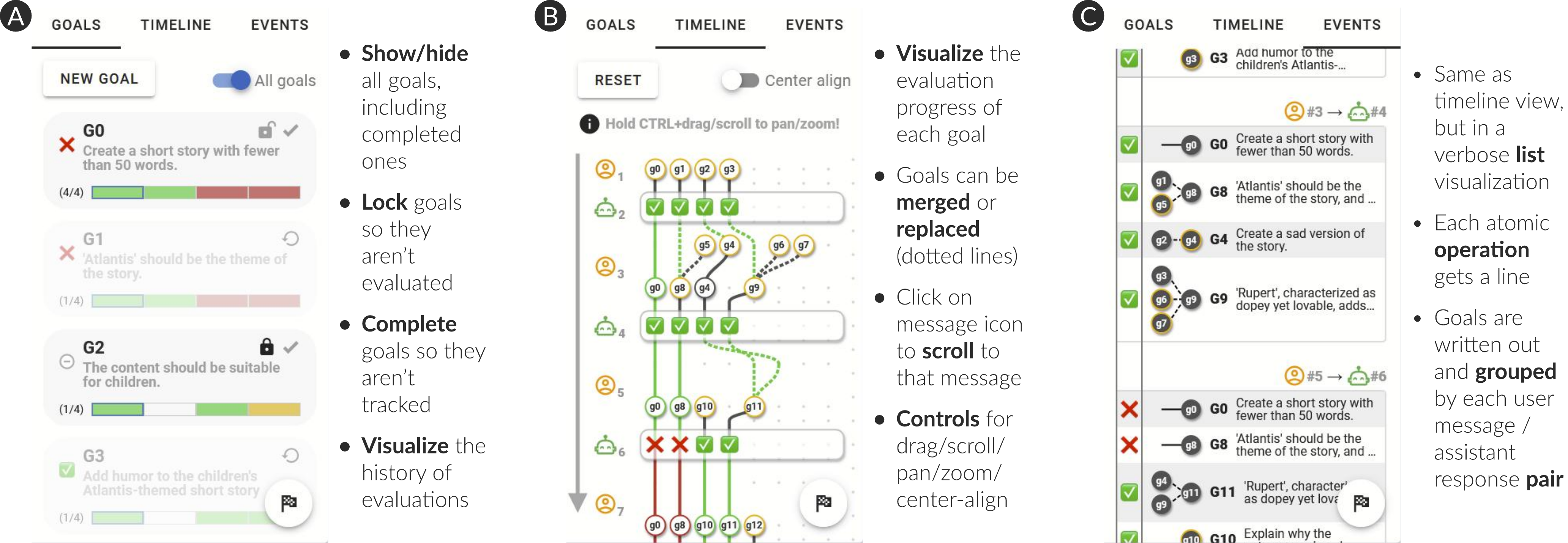}
  \caption{%
    The three tabs in the progress panel.
    In the goals tab (\textbf{A}), users can control goals by locking or completing them, create their own goals, and visualize a goal's evaluation history.
    In the timeline tab (\textbf{B}), the goal pipeline is visualized as a node-link diagram, where nodes flow top to bottom and are colored by their evaluation.
    In the events tab (\textbf{C}), the goal pipeline operations are visualized in a list view with more verbose descriptions.
    Users can brush and link a row in the timeline or events view with its corresponding message in the chat by clicking the numbered user icons.
  }%
  \Description{%
    Three sub-figures labeled A, B, and C.
    Each sub-figure shows a screenshot of the side bar on the right of the OnGoal web interface.
    A shows Goals tab, B shows the Timeline tab, and C shows the Events tab.
  }%
  \label{fig:progress_view}
\end{figure*}
}{
}

\aptLtoX[graphic=no, type=html]{
}{
\begin{figure*}[!t]
  \centering
  \includegraphics[width=\linewidth]{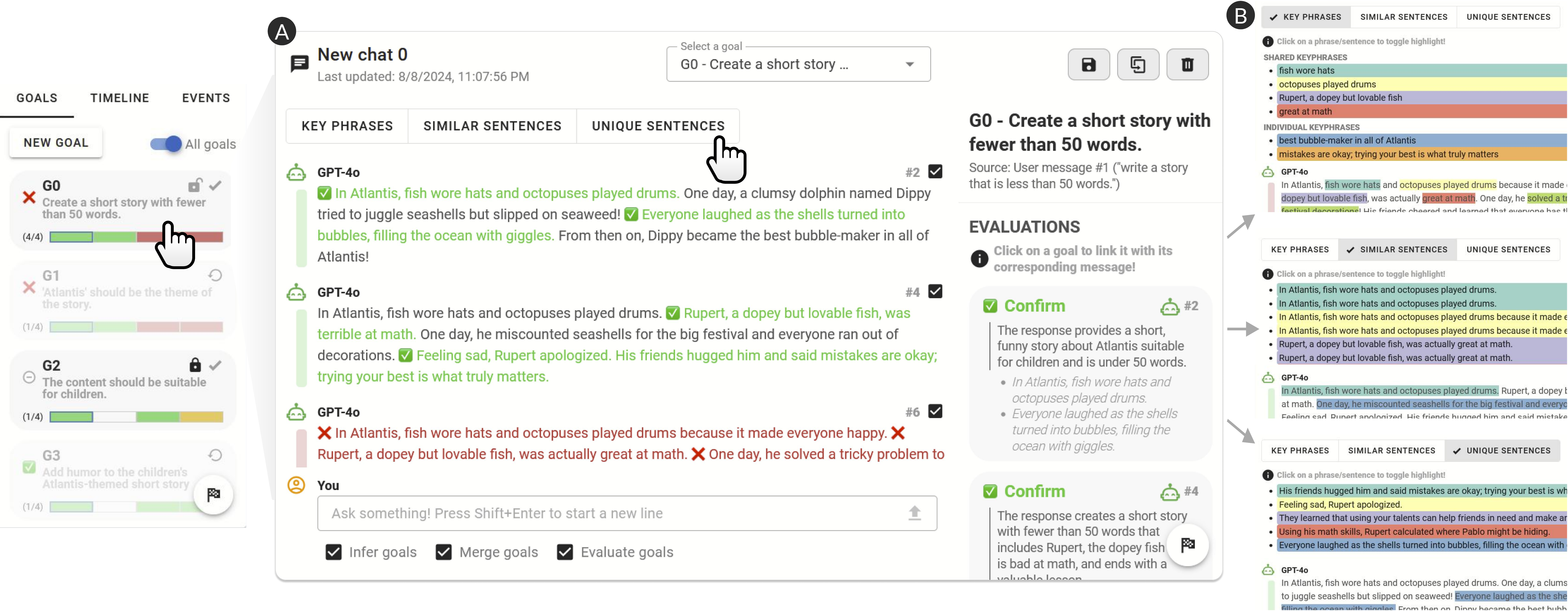}
  \caption{%
    Clicking on a goal in the goals tab (Fig.~\ref{fig:progress_view}) filters the chat by LLM responses that were evaluated against that goal (\textbf{A}).
    The evaluations and examples are highlighted in the text by default; clicking an evaluation in the side panel scrolls the conversation to that message.
    Three text highlighting modes (key phrases, similar and unique sentences) can be toggled to overlay on the chat log (\textbf{B}).
  }%
  \Description{%
    Two sub-figures labeled A and B.
    A is in the center and shows a filtered chat interface as the individual goal view.
    To the left of A is a smaller screenshot of the Goals tab with a cursor clicking a widget of a goal, indicating the operation to access the individual goals view.
    B is to the right of A and shows three sub-sub-figures of A stacked vertically.
    From top to bottom, the sub-sub-figures show text highlighting of key phrases, similar sentences, and unique sentences in the individual goals view.
  }%
  \label{fig:individual_view}
\end{figure*}
}

\subsubsection{Ex-situ progress panel}
\label{sec:exsitu_progress}
Next to the chat, the \textbf{progress panel} (Fig.~\ref{fig:teaser}B) allows users to track and control specific goals across three tabs (\textbf{C2}).
Each tab achieves different evaluation/review objectives: controlling the goal pipeline, visualizing the evaluation and merging of goals over time, and listing all pipeline events for validation.

The first tab is the \textbf{goals tab} (Fig.~\ref{fig:progress_view}A), which lists all final goals from the pipeline as widgets with a text description and several controls for the goal pipeline.
Users can \textit{lock} goals from being merged, \textit{complete} goals to remove them from evaluation, and \textit{restore} previously merged goals.
The second tab, the \textbf{timeline tab} (Fig.~\ref{fig:progress_view}B), is a Sankey-based \cite{Tufte:1983:VisualDisplayInfo} timeline visualization that summarizes the history of infer, merge and evaluation events from the goal pipeline.
The plot flows from top to bottom in sets of three rows for each user prompt / LLM response pair: row $1$ shows inferred goals and draws lines from above connecting to existing goals; row $2$ lists the final goals and draws lines from the existing and inferred goals above to represent combine, replace, and keep operations; and row $3$ shows icons representing the evaluation result from the final goal in the row above: a check for confirm, a cross for contradict, and a prohibited ``no'' sign for ignore.
The third tab is the \textbf{events tab} (Fig.~\ref{fig:progress_view}C), which displays events of the goal pipeline in a list, grouped by each user prompt / LLM response pair.
Within the timeline and events tabs, users can click on the numbered user or LLM icons to have the chat window scroll to the corresponding message.

\subsubsection{Individual goal view}
\label{sec:goal_view}
Users can drill down to get details of an individual goal on demand by selecting it in the goals tab.
This filters the chat window to only show LLM responses which were evaluated against the selected goal (Fig.~\ref{fig:individual_view}A), facilitating easier comparison of LLM responses over time (\textbf{C2}).
For example, by scanning all messages related to a single goal, users can identify repeating phrases, forgotten requests, or inconsistent responses.
In this view, the \textbf{progress panel} transforms to show a list of the evaluations for that goal (Fig. \ref{fig:usage}C); selecting an evaluation scrolls the chat window to the corresponding LLM response.

\subsection{Highlighting Text to Support Sensemaking}
\label{sec:text_highlighting}

\aptLtoX[graphic=no, type=html]{
\begin{figure*}[!t]
  \centering
  \includegraphics[width=\linewidth]{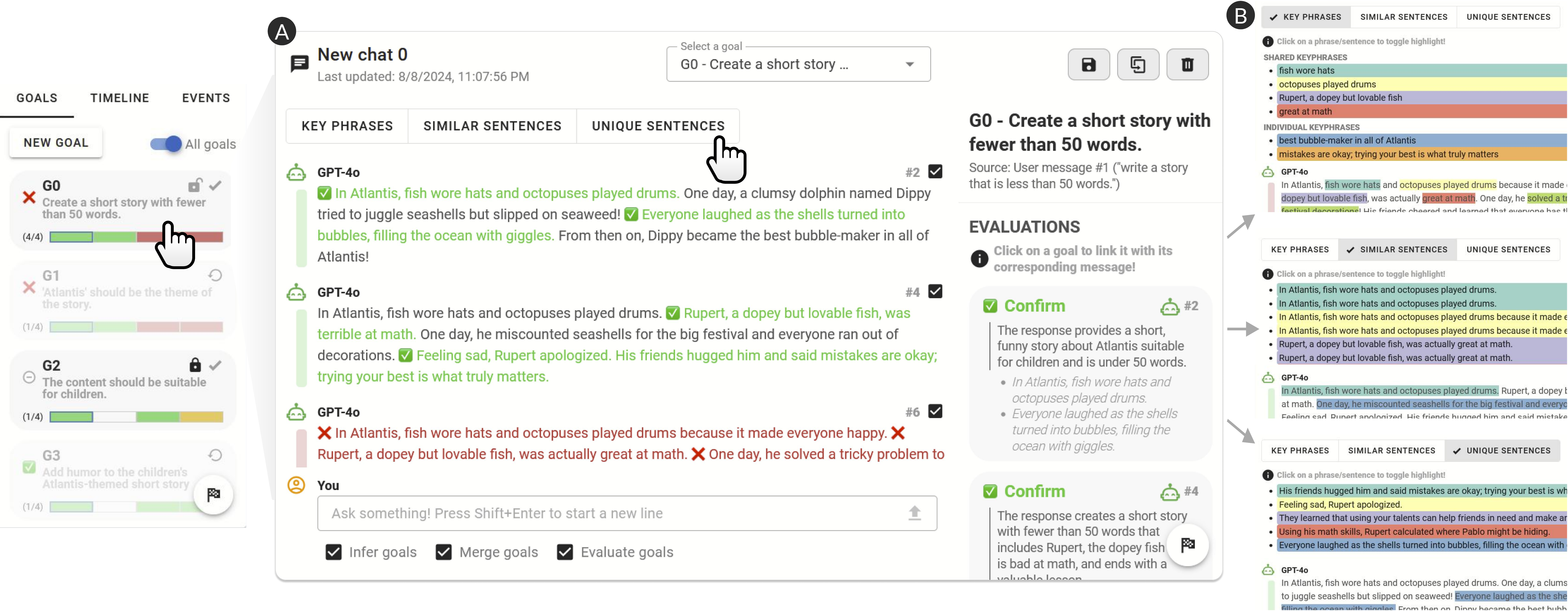}
  \caption{%
    Clicking on a goal in the goals tab (Fig.~\ref{fig:progress_view}) filters the chat by LLM responses that were evaluated against that goal (\textbf{A}).
    The evaluations and examples are highlighted in the text by default; clicking an evaluation in the side panel scrolls the conversation to that message.
    Three text highlighting modes (key phrases, similar and unique sentences) can be toggled to overlay on the chat log (\textbf{B}).
  }%
  \Description{%
    Two sub-figures labeled A and B.
    A is in the center and shows a filtered chat interface as the individual goal view.
    To the left of A is a smaller screenshot of the Goals tab with a cursor clicking a widget of a goal, indicating the operation to access the individual goals view.
    B is to the right of A and shows three sub-sub-figures of A stacked vertically.
    From top to bottom, the sub-sub-figures show text highlighting of key phrases, similar sentences, and unique sentences in the individual goals view.
  }%
  \label{fig:individual_view}
\end{figure*}
}{
}

Reading lengthy LLM responses as plain text can make it challenging to identify patterns about LLM behaviors.
Gero et al. show that text highlighting can improve sensemaking behaviors and insights when evaluating LLM responses at scale \cite{Gero:2024:SensemakingLLMsAtScale}.
Drawing inspiration from their insights, OnGoal presents two variations of text highlighting to help users quickly identify relevant clauses in the text related to potentially problematic LLM behaviors (\textbf{C3}).

First, goal evaluations include examples extracted verbatim from the LLM response that are highlighted inline when a goal explanation panel is open (Fig.~\ref{fig:individual_view}A).
\add{Examples are colored by evaluation result: green for examples that show a goal is confirmed, yellow for ignored, and red for contradicted.}
The highlighted clauses exemplify local LLM behaviors that are potentially related to how and why it evaluated a particular goal based on the prior user message.
For example, users can compare highlighted texts across multiple responses (\textbf{C3}) to assess the consistency of an LLM response regarding specific goals or determine if progress is being made.

Second, in the individual goal view, the same goal evaluation examples are highlighted in every LLM response by default.
Adapting a subset of techniques from Gero et al. \cite{Gero:2024:SensemakingLLMsAtScale}, OnGoal lets users toggle between three additional text highlighting techniques (Fig.~\ref{fig:individual_view}B): (1) using a generative LLM and prompt engineering, OnGoal extracts all \textit{key phrases} from each LLM response and highlights the shared and unique phrases; OnGoal tokenizes all sentences, computes a text embedding using a generative LLM, and computes pairwise cosine similarity between sentences, revealing (2) sentence pairs with the highest similarity (i.e. \textit{similar sentences}) and (3) the sentences with the lowest average similarity to all other sentences (i.e., \textit{unique sentences}).
These techniques can enable comparison of LLM response at scale related to an individual goal, potentially revealing insights related to global LLM behaviors including distractions and topic drift over time (\textbf{C3}).
The prompt for extracting key phrases is included in the \add{appendix (Sect.~\ref{sec:prompt_keyphrase})}\strike{ supplemental material}.

\subsection{Usage Scenario}
\label{sec:use_case}

\begin{figure*}[!t]
  \centering
  \includegraphics[width=\linewidth]{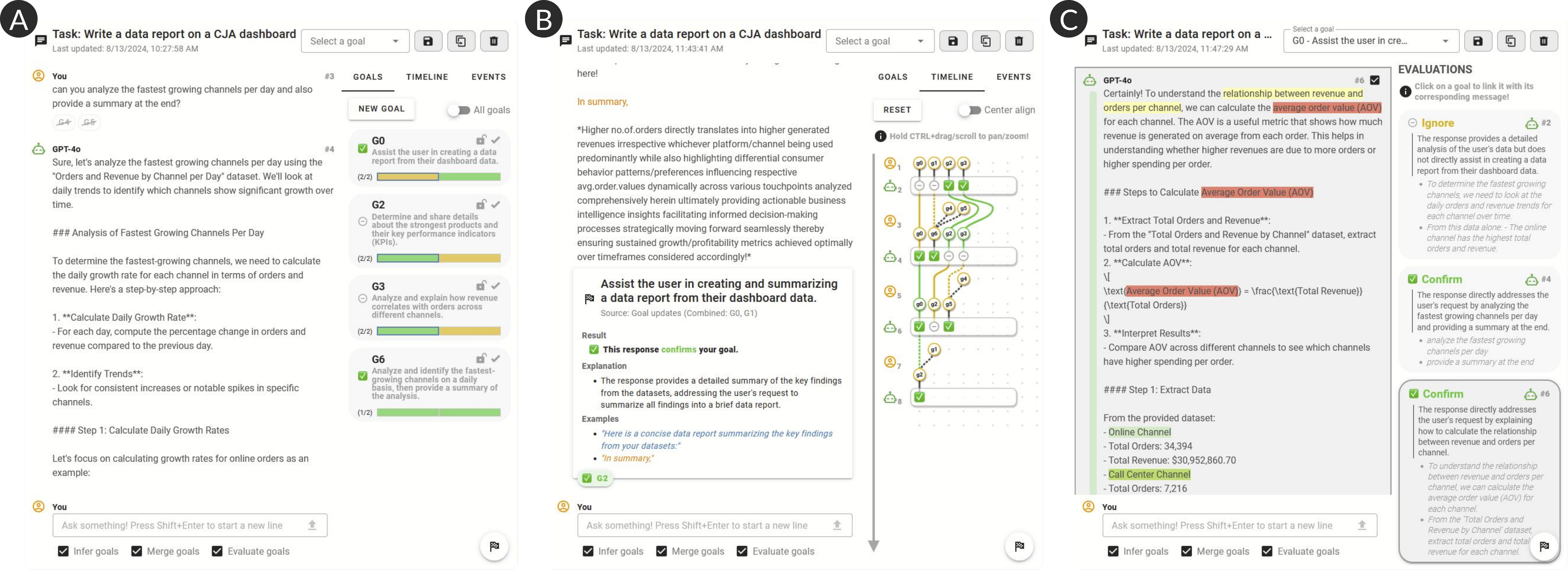}
  \caption{%
    Usage scenario of using OnGoal to analyze a CSV dataset.
    After giving the LLM the dataset, the user iteratively forms goals, explores and refines them, evaluates the explanations and examples, and reviews progress over time, using the features of OnGoal to support their workflow.
  }%
  \Description{%
    Three sub-figures labeled A, B, and C.
    Each sub-figure shows a screenshot of the entire OnGoal web interface.
    A shows the chat interface with the Goal tab open, B shows a goal glyph open with the Timeline tab open, and C shows the individual goal view with goal G0 and message number 6 selected.
  }%
  \label{fig:usage}
\end{figure*}

To illustrate how OnGoal can be used to support tasks accomplished via multi-turn dialogue with an LLM, we demonstrate a usage scenario involving exploratory data analysis (Fig.~\ref{fig:usage}).
Consider Jim, a sales associate working with a customer analytics dataset.
With limited time, Jim seeks assistance from an LLM chatbot to quickly extract insights from the data and use them to write a data report.
Jim turns to OnGoal to help manage the conversation.

Jim begins by uploading a CSV dataset and asking initial questions (Fig.~\ref{fig:usage}A), including basic questions like \textit{``What are the fastest growing market segments?''} as well as more advanced ones like \textit{``What is the relationship between revenue and rating per age group?''} 
OnGoal immediately provides goal glyphs under the LLM responses, allowing Jim to evaluate how each of the analysis goals was satisfied with explanations.
Jim noticed two of goals are being ignored --- the high-level goal of writing a data report, and the goal about information on the fastest growing market segments.
Recognizing that the goals are ignored, Jim is encouraged to dive deeper into the answers and ask follow ups to get more information.

Jim decides to address the ignored goals one by one.
This process helps clarify the intention behind the conversational goals, both for the LLM, and for Jim.
For example, Jim asks a clarifying question about the market growth rate, and gets more details that address what the assistant did not touch on before (Fig.~\ref{fig:usage}B).
The goal glyphs and goal tab in the progress view both show progress in confirming goals over time.
As Jim continues to chat, they learn more about their goals and update them as new information becomes available.
Jim uses the timeline tab in the progress view to reflect on their progress and how their goals have changed over time.
The timeline also shows progress, as less goals are carried over each turn.
When the assistant struggles, Jim clicks on a goal in the goal tab and dives deeper into the history of messages for that goal.
They analyze the similarities and differences between responses using key phrases to see if there is any missing or contradicting information (Fig.~\ref{fig:usage}C).
For example, by reviewing similar sentences, Jim observes the assistant repeating old information without updating based on new evidence.

Overall, this iterative process helped Jim better understand how to communicate their goals to the assistant, reduced the effort in remembering and tracking their goals, and helped them keep the assistant aligned with their goals over time.

\subsection{Implementation}
\label{sec:implementation}
The goal pipeline is implemented in Python using the OpenAI library to make API calls to GPT-4o and stream the results to the frontend interface.
OnGoal can work with any decoder LLM architecture and easily be adapted to use open-source LLMs as well as local LLMs, e.g., using PyTorch.
The OnGoal frontend is implemented in Vue.js; the visualizations are built using D3.js \cite{Bostock:2011:D3}.

\section{User Study}
\label{sec:evaluation}

Our goal for evaluation was to assess how OnGoal helps users address challenges of multi-turn dialogue with LLMs when managing conversational goals.
To do this, we conducted a between-subjects study, observing some participants using OnGoal to complete a writing task while other participants used a baseline interface without goal tracking and visualization for the same task.
\add{We focused on writing tasks to provide a controlled, yet realistic scenario for multi-turn dialogue with LLMs, which also allowed us to better tune the goal inference prompts for our specific use case.}
We compared usability issues in the baseline condition with those uncovered in OnGoal, paying close attention to which issues observed in the baseline condition are mitigated with OnGoal and how users utilized the features in OnGoal to evaluate and review their goals.

\aptLtoX[graphic=no, type=html]{
}{
\begin{table*}[!t]
    \caption{%
      Six conversational goals that were used in the study task.
    }%
    \Description{%
      Table with three columns and four rows.
      The first row is a header row.
      The first column in the header row has no label, the second is labeled "Editor \#1: The Creative Boss", and the third is labeled "Editor \#2: The Pragmatic Boss".
      The last three rows of the first column are labels for the rows to the right; from top to bottom, Language Style, Persuasiveness, and Imagery and Metaphors.
      The last three rows of the second and third column are the data.
    }%
    \label{tab:study_conversational_goals}
    \begin{tabular}{@{}rlll@{}}
        \toprule
        & \textbf{Editor \#1: The Creative Boss} & \textbf{Editor \#2: The Pragmatic Boss} & \\
        \midrule
        \textbf{Language Style} & 1. Use non-formal, conversational language & 2. Use formal and technical language & \\
        \textbf{Persuasiveness} & 3. Engage storytelling and emotional appeal  & 4. Build credibility through research and evidence   & \\
        \textbf{Imagery and Metaphors} & 5. Include rich imagery and creative metaphors & 6. Prefer facts over figurative language  & \\
        \bottomrule
    \end{tabular}
\end{table*}
}

\subsection{Study Design}
\label{sec:study_design}
We designed a \add{$1\times2$}\strike{ $2\times2$} between-subjects study with $20$ participants.
Our study compares user performance and behavior across two interface conditions: a baseline interface presenting a standard LLM-based chat experience with limited goal tracking and no visualization features; and the full OnGoal interface.
Each participant was assigned one of two\add{ similar} writing tasks\add{ to complete, limiting potential confounds from learning and task effects. The differences in these tasks were not compared}.

\medskip
\noindent\textbf{Interfaces. }
As part of the design rationale (Sect.~\ref{sec:methodology}), we observed that many common issues with multi-turn dialogue could be related to interface design.
To understand if OnGoal could address any of these issues, we aimed to gain a baseline measure of user behaviors when using a typical LLM chat interface and measure the difference in performance with OnGoal.
To ensure consistency in the goal inference pipeline and LLM responses between interfaces, we utilized the same LLM framework as OnGoal.

We set up the baseline interface as a barebones LLM chat interface using the same framework as OnGoal but stripped of any visualizations and with merging and evaluation disabled.
Because our task was to address fixed conversational goals (described under \textbf{Tasks}), we reconfigured the progress panel to show a limited view of the goals tab with only goal inference enabled in the pipeline.
In other words, baseline users would see the initial list of goals and goals being inferred after each conversational turn, but no merging or evaluation in the progress panel.
In both conditions, we preloaded the fixed goals in both interfaces and locked them.
This ensured both conditions had equal ability to see the same goals, but only OnGoal would have access to visualizations and the results of merge and evaluation.
To further control for user behaviors, we also disabled the new goal button, the goal controls (lock, complete, restore), and the goal pipeline checkboxes in the OnGoal condition, as these were not related to the task.

\aptLtoX[graphic=no, type=html]{
\begin{table*}[!t]
    \caption{%
      Six conversational goals that were used in the study task.
    }%
    \Description{%
      Table with three columns and four rows.
      The first row is a header row.
      The first column in the header row has no label, the second is labeled "Editor \#1: The Creative Boss", and the third is labeled "Editor \#2: The Pragmatic Boss".
      The last three rows of the first column are labels for the rows to the right; from top to bottom, Language Style, Persuasiveness, and Imagery and Metaphors.
      The last three rows of the second and third column are the data.
    }%
    \label{tab:study_conversational_goals}
    \begin{tabular}{@{}rlll@{}}
        \toprule
        & \textbf{Editor \#1: The Creative Boss} & \textbf{Editor \#2: The Pragmatic Boss} & \\
        \midrule
        \textbf{Language Style} & 1. Use non-formal, conversational language & 2. Use formal and technical language & \\
        \textbf{Persuasiveness} & 3. Engage storytelling and emotional appeal  & 4. Build credibility through research and evidence   & \\
        \textbf{Imagery and Metaphors} & 5. Include rich imagery and creative metaphors & 6. Prefer facts over figurative language  & \\
        \bottomrule
    \end{tabular}
\end{table*}
}{
}

\medskip
\noindent\textbf{Tasks. }
We designed two different writing tasks for our study\add{ from the following template --- ``As a writer working for two bosses, use an LLM to draft a single article that satisfies the goals of both bosses.''}
The \add{following instructions were}\strike{ task was} shown to participants\add{, with either 1. or 2. substituted to counterbalance across participants}:

\begin{quote}
    \textit{As a writer for an online blogging company, use GPT to write a single, five-paragraph article of tips on \{\textbf{1.} traveling to a destination of your choice for a weekend, \textbf{2.} hosting an event of your choice at your residence\}. You work for two different bosses, each with three specific goals for how the article should be written. You must satisfy all of their goals to the best of your ability.}
\end{quote}

\noindent
Participants were shown the set of six goals (Table~\ref{tab:study_conversational_goals}), three from each boss reflecting their personal expectations on style and content of the article.
These goals are not inferred from the goal pipeline, but instead preloaded and fixed in both interfaces.
This way, all participants had an equal opportunity to see and track them using the progress panel and, in the OnGoal condition, use the evaluation stage of the pipeline to evaluate the goals.
While these specific goals were not part of the goal pipeline, other goals could still be inferred, merged, and evaluated in the OnGoal interface.
Participants were required to satisfy all goals as much as possible and to use the LLM to generate all text included in the final article, but were allowed to use the LLM chat interface in whatever way they wanted to (e.g., copy-pasting a template, writing custom instructions, etc.).

Writing tasks are commonly used as a task for evaluating LLM interfaces \cite{Arawjo:2024:ChainForge, Tongshuang:2022:PromptChainer}.
Writing typically involves satisfying \add{global} goals such as \textit{``use a positive tone''}, which our system \add{was designed to}\strike{ can} interpret as a conversational goal to guide the LLM response\add{ (Sect.~\ref{sec:goal_pipeline})}.
We aimed to present goals that were sometimes contradictory, requiring users and the LLM to make intentional decisions on which goals to address, how to address them, and why they addressed them.
\add{We discussed our task design with professional copy editors, who confirmed that having multiple bosses with conflicting writing goals for a single article often happens, requiring them to balance job requirements with self-evaluation. This scenario is also useful to study how users manage trade-offs between addressing LLM issues and making decisions, and how this affects human-AI alignment.}\strike{ Thus, the overarching objective of our writing tasks was to encourage users to engage in a multi-turn dialogue rather than single-shot prompting to accomplish multiple conversational goals over time. We counterbalanced two tasks to overcome potential task bias in our evaluation.}

\add{In addition to writing the article, participants were asked to evaluate and review their goals after each message they sent. Upon getting an LLM response, both interfaces present}\strike{ After each prompt that a participant sent to the LLM, the participant would be asked} two sets of two \textbf{initial} questions (labeled `I' in Fig.~\ref{fig:performance}, \ref{fig:interactions}), rated on a scale from $1-5$.
The first question in each set\add{, broken into two parts,} was: (1) to \textbf{evaluate} whether the LLM's response satisfied a single goal chosen at random; and (2) to \textbf{review} whether that same goal chosen at random had been consistently satisfied or dissatisfied across all messages.
The second question in each set\add{, presented after each part,} asked the participant how confident they were in their assessment.
For example, for the goal \textit{``2. Use formal and technical language''}, the participant would evaluate if the LLM's current response used formal and technical language ($1-5$), if all of the LLM's prior responses also used formal and technical language ($1-5$), and how confident they were in \add{evaluating ($1-5$) and reviewing ($1-5$)}\strike{ their answers ($1-5$)}.
The order of the evaluate/review questions in each set was randomized to counterbalance learning effects in tracking time spent.
After the task was completed, the participant's previous answers were then shown back to them with the interface reloaded to how it appeared at the time when they answered.
They would then have an opportunity to \textbf{validate} (labeled `V' in Fig.~\ref{fig:performance}, \ref{fig:interactions}) and revise their answer if they wanted to, using any features of the interface to help them, without the time pressure or the task to worry about.

By explicitly asking participants to evaluate and review their goals, we aimed to elicit think-aloud feedback on how the interfaces either did or did not support these tasks, as well as usability findings in the post-study survey and interview related to their overall experience evaluating and reviewing conversational goals.
Further, by giving participants the opportunity to revise their answers after the task, we could calibrate findings on time spent, effort expended, and confidence with how participants felt about their initial answer during the task.
We analyzed the self-reported answers and report differences in task performance in our usability findings (Sect.~\ref{sec:usability_findings}).

\medskip
\noindent\textbf{Procedure. }
Each study lasted 60 minutes in total.
We employed a between-subjects design, where participants used one of the two possible interfaces and tasks ($2\times2$), chosen at random to control for ordering effects, resulting in $10$ participants using our baseline interface (P$1-10$) and $10$ using OnGoal (P$11-20$).
Between-subjects allowed us to compare independently sampled distributions of self-reported performance factors including time spent, effort, and confidence between interfaces in our analysis, while ensuring participants using one interface did not know about the other.

After giving consent, participants first self-reported their demographics in a pre-study survey.
Then, the researcher conducting the session explained aloud the various features of the chosen interface and the expectations of the task.
Participants then performed a practice task for 12 minutes in the same style as the live task that was not recorded.
After, participants spent 15 minutes completing the live task which was recorded.
In both the practice and live task, participants were instructed to follow a think-aloud protocol \cite{Ericsson:1984:ProtocolAnalysis}, explaining their thoughts, insights, and feedback as they worked.
Finally, after they completed both tasks, the participant completed a post-study usability survey, and both the participant and researcher engaged in a semi-structured debrief interview discussing the participant's experience during the session.

\aptLtoX[graphic=no, type=html]{
}{
\begin{figure*}[!t]
\centering
\includegraphics[width=\linewidth]{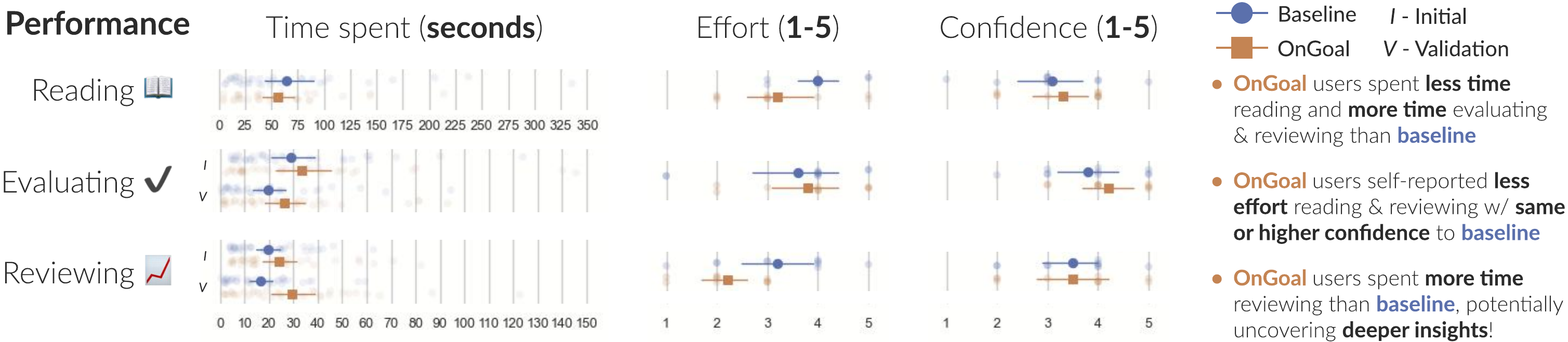}
  \caption{%
    Time spent reading messages, evaluating goals, and reviewing goals between interfaces and task phases.\strike{ Less time spent reading and more time spent evaluating and reviewing with OnGoal may indicate users are prioritizing managing goals over exhaustive reading. The gap between interfaces for reviewing specifically is much higher when users are not pressed for time in validation, potentially indicating they are using the time to learn more about how the LLM is working to better communicate their goals.}
  }%
  \Description{%
    Three columns and three rows of horizontal point plots with 95\% confidence intervals.
    The columns are time spent in seconds, effort from 1 to 5, and confidence from 1 to 5.
    The rows are reading, evaluating, and reviewing.
    Each point plot has two rows, one for baseline and one for OnGoal.
    The range of the point plot for time spent reading is 0 to 350, while the range of the point plots for time spent evaluating and reviewing are 0 to 150.
  }%
  \label{fig:performance}
\end{figure*}
}

\aptLtoX[graphic=no, type=html]{
}{
\begin{figure}[!t]
\centering
\includegraphics[width=\linewidth]{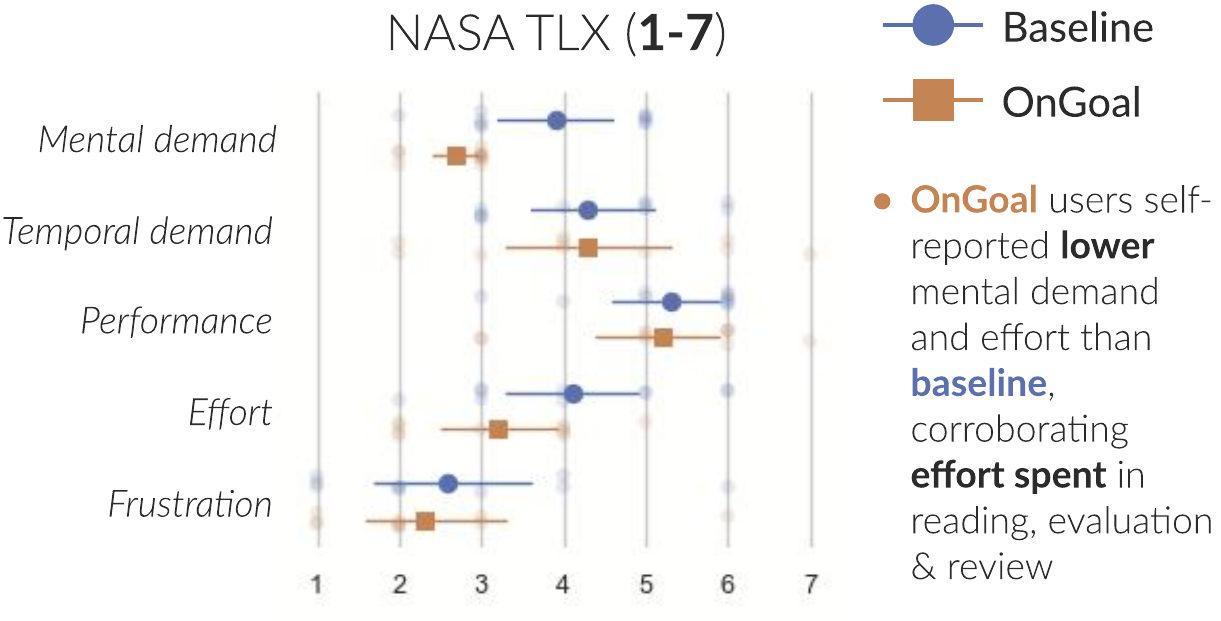}
  \caption{%
    Responses to the NASA Task Load Index (TLX) for assessing workload between interfaces.\strike{ Mental demand and effort of the task are significantly lower for OnGoal than baseline, corroborating self-reported effort in Fig.~\ref{fig:performance}.}
  }%
  \Description{%
    One horizontal point plot with 95\% confidence intervals and a range from 1 to 7.
    The title of the point plot is NASA TLX (1-7).
    The point plot has five rows labeled Mental demand, Temporal demand, Performance, Effort, and Frustration.
    Each row has two sub-rows, one for baseline and one for OnGoal.
  }%
  \label{fig:tlx}
\end{figure}
}

\aptLtoX[graphic=no, type=html]{
}{
\begin{figure*}[!t]
\centering
\includegraphics[width=\linewidth]{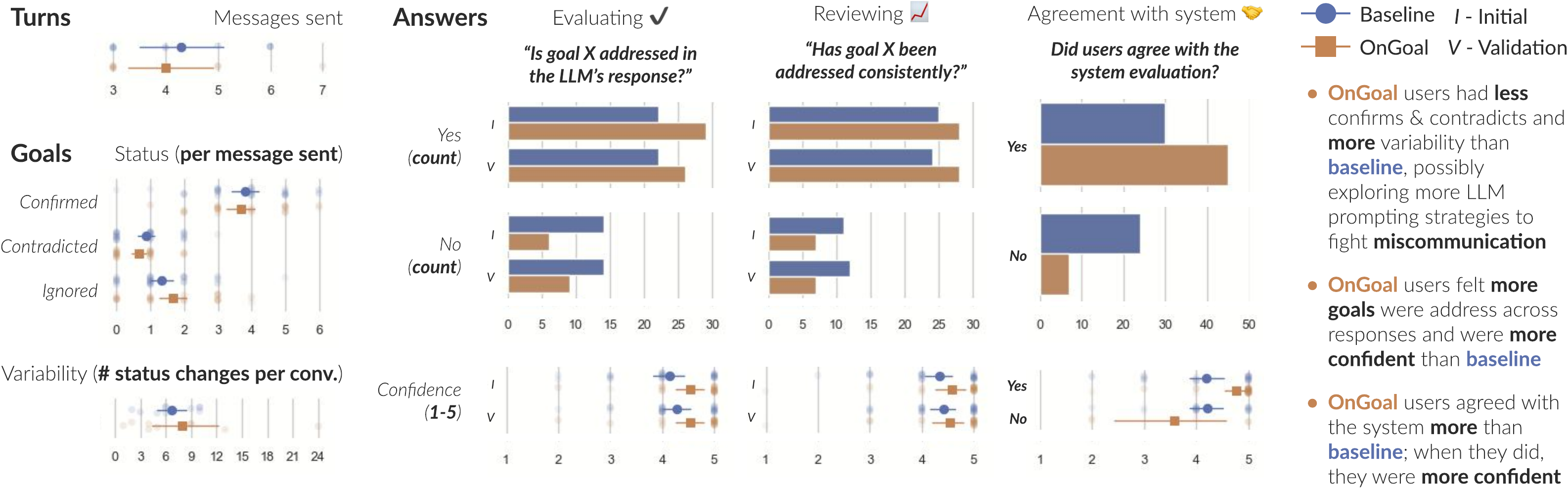}
  \caption{%
    Turns taken, goals confirmed / contradicted / ignored, goal variability, and final answers to the initial and validation questions between interfaces.
    \add{Answers in the evaluating and reviewing columns were recorded as questions participants responded to after each dialogue turn (Sect.~\ref{sec:study_design}, \textbf{Tasks}). We measured agreement between participant answers and the system's evaluations post-hoc.}\strike{ Less goals confirmed and contradicted, with more ignored, may suggest that OnGoal users experimented with how they structured their goal communication to get different results. The higher variability of goal status changes with OnGoal may further corroborate this. OnGoal users felt more goals were addressed successfully and were more confident in their answers, particularly when they agreed with the system evaluation.}
  }%
  \Description{%
    Three sub-figures of horizontal point plots with 95\% confidence intervals and horizontal bar charts.
    All point plots and bar charts have two sub-rows, one for baseline and one for OnGoal.
    On the top-left, a header labeled Turns and a single point plot with range 3 to 7 labeled Messages sent.
    On the bottom-left, two point plots with the same header labeled Goals.
    The top point plot is labeled Status (per message sent) with a range of 0 to 6 and has three rows labeled Confirmed, Contradicted, and Ignored.
    The bottom point plot is labeled Variability (\# status changes per conv.) with a range of 0 to 24.
    On the right, 6 bar charts and 3 point plots under a header labeled Answers.
    The bar charts in a three column by two row configuration.
    The rows are Yes (count) and No (count).
    The columns are Evaluating with a range 0 to 30, Reviewing with a range 0 to 30, and Agreement with system with a range 0 to 50.
    Below the bar charts are three point plots with range 1 to 5, aligned horizontally with the same columns as the bar charts.
    The single row is labeled Confidence (1-5).
  }%
  \label{fig:interactions}
\end{figure*}
}

\aptLtoX[graphic=no, type=html]{
}{
\begin{figure}[!t]
\centering
\includegraphics[width=\linewidth]{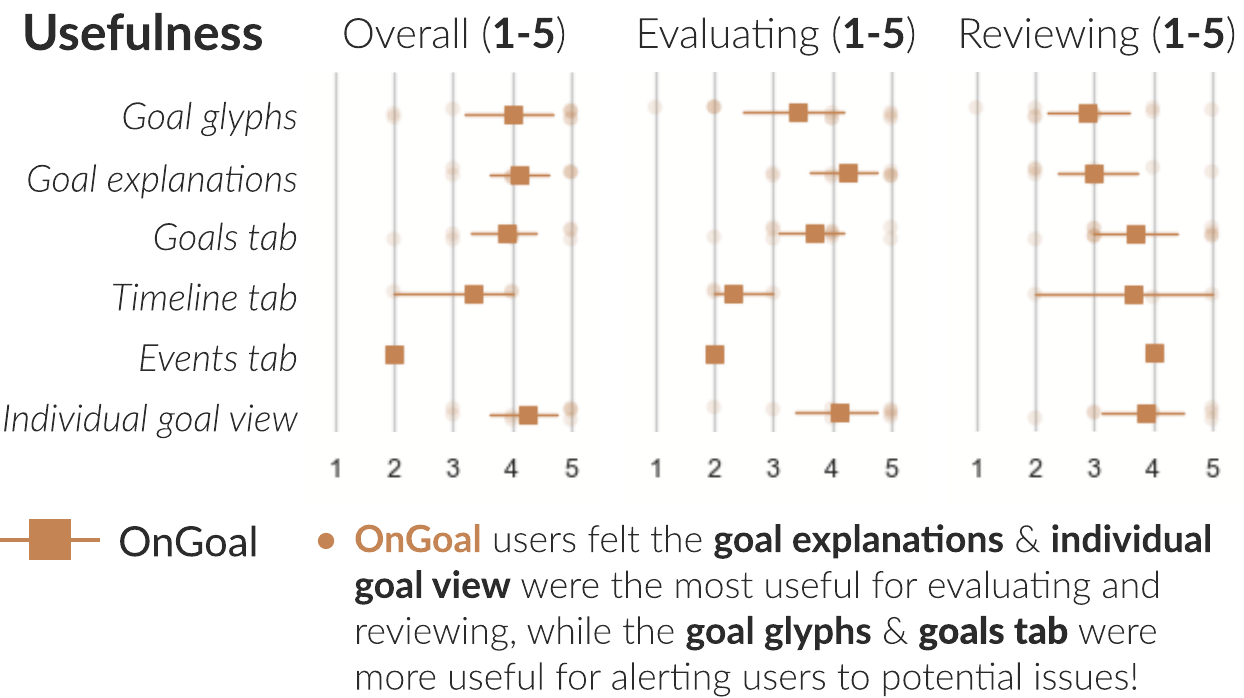}
  \caption{%
    Usefulness ratings for various OnGoal features.\strike{ The goal explanations and the individual goal view helped users to identify problematic LLM behaviors, while the goal glyphs and the tabs in the progress panel helped alert users to changes in their goal progress.}
  }%
  \Description{%
    Three side-by-side sub-figures of horizontal point plots with 95\% confidence intervals, each with a range 1 to 5.
    All point plots share six rows labeled Goal glyphs, Goal explanations, Goals tab, Timeline tab, Events tab, and Individual goal view, as well as one sub-row for OnGoal for each main row.
    The left sub-figure is labeled Overall (1-5), the middle sub-figure is labeled Evaluating (1-5), and the right sub-figure is labeled Reviewing (1-5).
  }%
  \label{fig:usefulness}
\end{figure}
}

\medskip
\noindent\textbf{Participants. }
We recruited $20$ participants (P$1-20$) at both an academic institution and private company via recruitment emails and interpersonal networking.
$14$ self-identified as male and $6$ self-identified as female, between the ages of $19$ and $37$, with a median age of $23$.
$16$ participants were pursuing higher education degrees in fields spanning Computer Science ($10$), Human-Computer Interaction ($4$), Human-Centered Computing ($1$), and Information Visualization ($1$).
Four participants held industry roles as researchers ($2$) and engineers ($2$).

All participants had prior experience using LLM chat interfaces, with all except one using them at least \textbf{once a month} or more.
Participants self-reported usage of LLM chat interfaces for various tasks; most used it for help with \textbf{programming} ($17$) and \textbf{writing} ($13$), some used it for \textbf{education} ($9$) such as learning topics, studying, and asking questions, or as a \textbf{personal assistant} ($5$), and a few used it for \textbf{research} ($6$) including summarizing papers, brainstorming ideas and analyzing data.
13 participants kept conversations short (i.e. less than five messages), 6 held medium-length conversations (between five and ten messages), and only 1 regularly had long conversations (more than 10 messages).

\medskip
\noindent\textbf{Measures and analysis. }
We asked participants to use a think-aloud protocol \cite{Ericsson:1984:ProtocolAnalysis} as they worked and summarize how they arrived at their results after completing tasks.
We recorded the screen and audio (participant and researcher) as well as interaction logs of all mouse events (clicks, hovers).
We also tracked users' performance in the task using their responses to the evaluation and review questions described in \textbf{Tasks} above.

We then analyzed participants' interactions and responses to the in-task questions and post-study surveys for insights on task performance.
Following Dragicevic \cite{Dragicevic:2016:HCIStats}, we generated and interpreted sample means as effect size using bootstrapped $95$\% confidence intervals (CIs) with $10,000$ resamples to represent uncertainty.
\add{For a given confidence level and sample size, CI width increases with increasing variability; results are considered significant if CIs do not overlap. We report all means and CIs for the baseline (B) and OnGoal (O) conditions, as well as the strength of the comparison, in the text as follows: ``(B: $mean$ $[CI]$, O: $mean$ $[CI]$, \{strong, weak, no\} evidence)'' We also visualize the raw data and $95$\% CI in each of our results figures (\ref{fig:performance}, \ref{fig:tlx}, \ref{fig:interactions}, \ref{fig:usefulness}) as point plots with error bars over top of the raw data as transparent dots.}
We further evaluated both the video recordings of each session and the audio recordings of participants' think-aloud protocol, debrief interview, and the researcher's notes, and conducted inductive thematic analysis \cite{Boyatzis:1998:ThematicAnalysis}, identifying emergent themes that were discussed amongst all authors relating to issues and workflows encountered using both interfaces.

\subsection{Study Results}
\label{sec:study_results}
Comparing survey responses, interaction logs, and think-aloud feedback across both tasks, key differences emerged in how participants used the baseline interface in comparison with OnGoal.

\subsubsection{Usability findings}
\label{sec:usability_findings}
We first analyzed participants' post-study survey results and interaction logs for \add{significant differences between interfaces. From the interaction logs, we analyzed time spent (in seconds), turns taken, goals addressed, and answers to the evaluation and review questions. From the post-study surveys, we analyzed self-reported levels of effort and confidence in completing tasks ($1-5$), the usefulness of OnGoal's features ($1-7$), and finally perceived accuracy of OnGoal's goal pipeline ($1-5$).}\strike{ key findings related to time spent, self-reported levels of effort and confidence completing tasks, interactions between interfaces including turns taken, addressing goals, and answers to the evaluation and review questions, and finally the usefulness of the OnGoal features.}

Overall, we observed \add{several significant differences (i.e., overlapping $95$\% confidence intervals)}\strike{ key differences} in usability between interfaces.
\add{With OnGoal, participants spent more time reviewing during validation, self-reported lower effort reading and reviewing, self-reported lower mental demand, and self-reported higher initial confidence in evaluating goals. We attribute many of these observations to OnGoal users }\strike{ OnGoal users spent less time reading and more evaluating and reviewing with similar confidence to baseline users. They potentially} changing their prompting strategy more often and generally feeling more in agreement with the LLM in terms of addressing goals than baseline users.
The most common use of OnGoal features was using the goal glyphs and progress panel to identify issues, goal explanations to evaluate how the LLM interpreted goals in their most recent message, and the individual goal view to review the history of messages for patterns related to addressing their goals over time.
\add{Finally, OnGoal users found the evaluation stage of the goal pipeline significantly less accurate than the infer, pointing to emergent issues in designing effective goal evaluations.}

\aptLtoX[graphic=no, type=html]{
\begin{figure*}[!t]
\centering
\includegraphics[width=\linewidth]{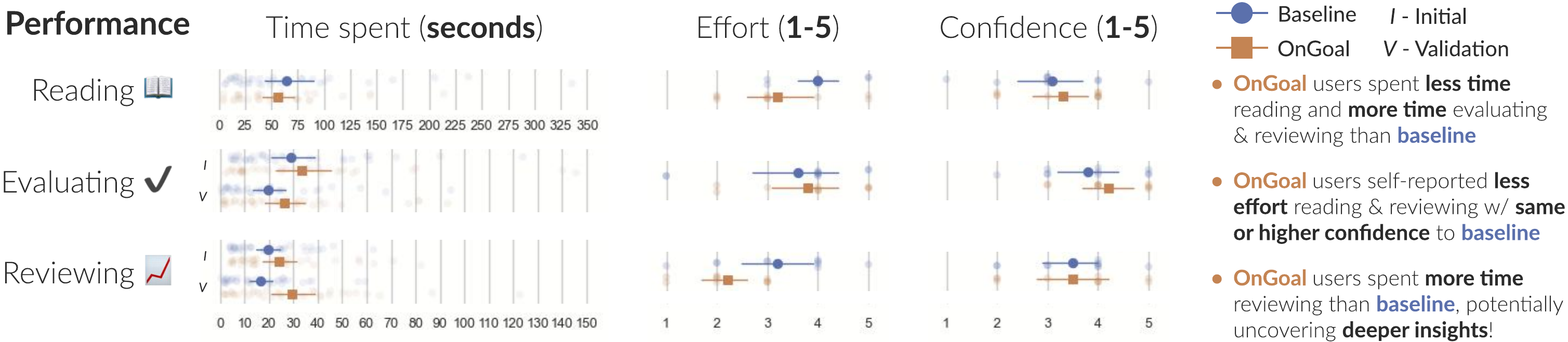}
  \caption{%
    Time spent reading messages, evaluating goals, and reviewing goals between interfaces and task phases.\strike{ Less time spent reading and more time spent evaluating and reviewing with OnGoal may indicate users are prioritizing managing goals over exhaustive reading. The gap between interfaces for reviewing specifically is much higher when users are not pressed for time in validation, potentially indicating they are using the time to learn more about how the LLM is working to better communicate their goals.}
  }%
  \Description{%
    Three columns and three rows of horizontal point plots with 95\% confidence intervals.
    The columns are time spent in seconds, effort from 1 to 5, and confidence from 1 to 5.
    The rows are reading, evaluating, and reviewing.
    Each point plot has two rows, one for baseline and one for OnGoal.
    The range of the point plot for time spent reading is 0 to 350, while the range of the point plots for time spent evaluating and reviewing are 0 to 150.
  }%
  \label{fig:performance}
\end{figure*}
}{
}

\aptLtoX[graphic=no, type=html]{
\begin{figure}[!t]
\centering
\includegraphics[width=\linewidth]{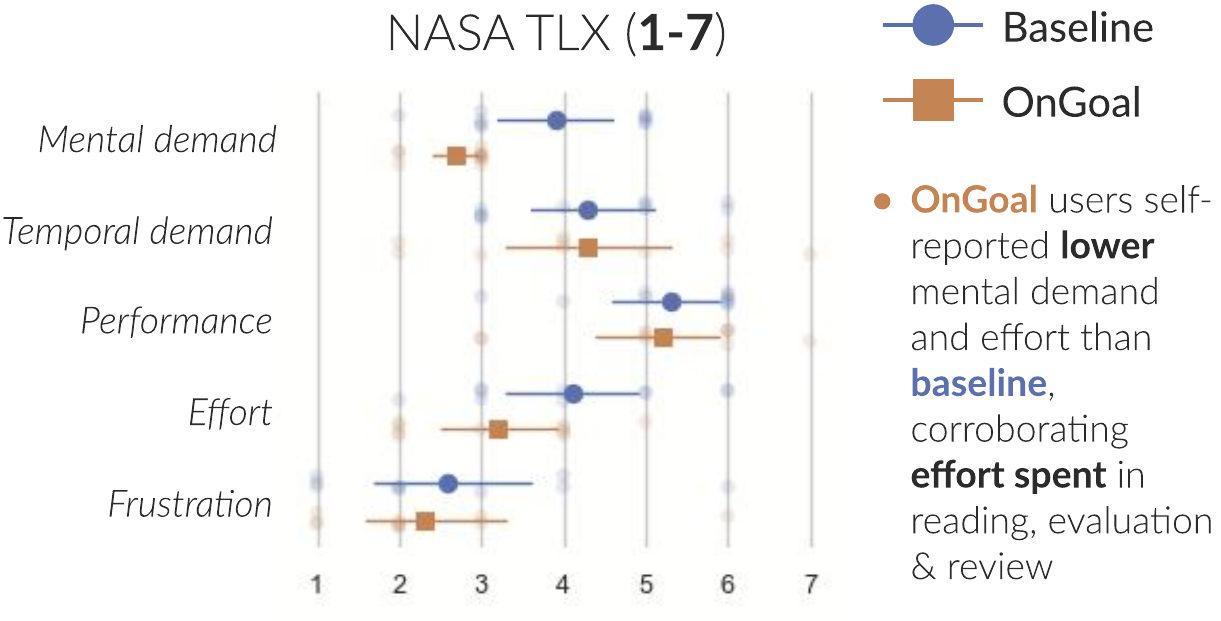}
  \caption{%
    Responses to the NASA Task Load Index (TLX) for assessing workload between interfaces.\strike{ Mental demand and effort of the task are significantly lower for OnGoal than baseline, corroborating self-reported effort in Fig.~\ref{fig:performance}.}
  }%
  \Description{%
    One horizontal point plot with 95\% confidence intervals and a range from 1 to 7.
    The title of the point plot is NASA TLX (1-7).
    The point plot has five rows labeled Mental demand, Temporal demand, Performance, Effort, and Frustration.
    Each row has two sub-rows, one for baseline and one for OnGoal.
  }%
  \label{fig:tlx}
\end{figure}
}{
}

\medskip
\noindent\textbf{Time, effort, and confidence. }
We conducted a post-study usability survey to get self-reported levels of effort and confidence in performing tasks and compared these reports to logs of time participants spent (Fig.~\ref{fig:performance}).

Compared with participants using the baseline interface, participants using OnGoal spent slightly more time evaluating\add{ (B: $29.8$ $[20.9, 38.6]$, O: $34.1$ $[22.9, 45.3]$, weak evidence)} and reviewing\add{ (B: $19.7$ $[15, 24.5]$, O: $24.4$ $[17.6, 31.2]$, weak evidence)} and less time reading\add{ (B: $66.5$ $[43.4, 89.6]$, O: $56.8$ $[42, 71.7]$, weak evidence)}.
The gap between time spent in the interfaces was slightly larger after the task ---validating evaluation took somewhat longer\add{ (B: $20.2$ $[13.5, 26.8]$, O: $26.6$ $[18.4, 34.9]$, weak evidence)}, while validating reviewing took \add{significantly} longer\add{ (B: $16.8$ $[12, 21.7]$, O: $30$ $[21.2, 38.9]$, strong evidence)}.
This could suggest participants are encouraged to spend more time reviewing their conversation with OnGoal when not pressed for time.

\add{This surprised us. Both conditions were encouraged to work on goal alignment, and both only had access to a text box to interact with the LLM with no suggestions on how to fix misalignment. With limited time, we expected OnGoal users to look for ways to offload cognition to the LLM to handle evaluating and reviewing, especially given that OnGoal was more information-heavy compared with baseline. Instead, OnGoal users actively engaged in evaluation and review more and in more ways; e.g., adapting their communication strategy to align with the LLM. Similar observations were reported in work on ``desirable difficulties'', where longer periods of reviewing and reflecting can enhance data understanding \cite{bjork2011making}. We discuss why these strategies only appear with OnGoal in Sect.~\ref{sec:thematic_analysis}.}

Further, participants self-reported lower levels of effort in reading\add{ (B: $4$ $[3.6, 4.4]$, O: $3.3$ $[2.6, 3.9]$, weak evidence)} and reviewing\add{ (B: $3.2$ $[2.5, 3.9]$, O: $2$ $[1.7, 2.6]$, strong evidence)}, corroborating lower self-reported mental demand\add{ (B: $3.9$ $[3.2, 4.6]$, O: $2.7$ $[2.4, 3.0]$, strong evidence)} and effort\add{ (B: $4.1$ $[3.3, 4.9]$, O: $3.2$ $[2.5, 3.9]$, weak evidence)} with similar performance in the NASA TLX (Fig.~\ref{fig:tlx}).
Confidence was reported similar for reading and evaluation, and slightly higher for evaluating with OnGoal\add{ (B: $3.8$ $[3.2, 4.4]$, O: $4.2$ $[3.7, 4.7]$, weak evidence)}.
Less effort may be needed to obtain similar results compared to a baseline interface.

\aptLtoX[graphic=no, type=html]{
\begin{figure*}[!t]
\centering
\includegraphics[width=\linewidth]{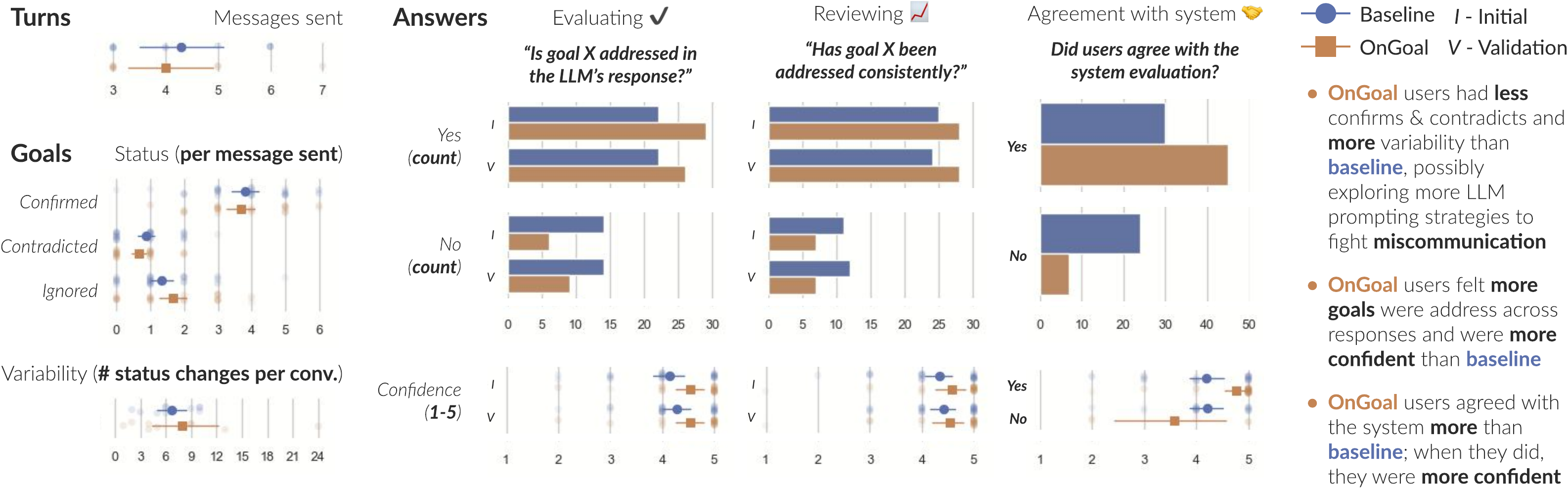}
  \caption{%
    Turns taken, goals confirmed / contradicted / ignored, goal variability, and final answers to the initial and validation questions between interfaces.
    \add{Answers in the evaluating and reviewing columns were recorded as questions participants responded to after each dialogue turn (Sect.~\ref{sec:study_design}, \textbf{Tasks}). We measured agreement between participant answers and the system's evaluations post-hoc.}\strike{ Less goals confirmed and contradicted, with more ignored, may suggest that OnGoal users experimented with how they structured their goal communication to get different results. The higher variability of goal status changes with OnGoal may further corroborate this. OnGoal users felt more goals were addressed successfully and were more confident in their answers, particularly when they agreed with the system evaluation.}
  }%
  \Description{%
    Three sub-figures of horizontal point plots with 95\% confidence intervals and horizontal bar charts.
    All point plots and bar charts have two sub-rows, one for baseline and one for OnGoal.
    On the top-left, a header labeled Turns and a single point plot with range 3 to 7 labeled Messages sent.
    On the bottom-left, two point plots with the same header labeled Goals.
    The top point plot is labeled Status (per message sent) with a range of 0 to 6 and has three rows labeled Confirmed, Contradicted, and Ignored.
    The bottom point plot is labeled Variability (\# status changes per conv.) with a range of 0 to 24.
    On the right, 6 bar charts and 3 point plots under a header labeled Answers.
    The bar charts in a three column by two row configuration.
    The rows are Yes (count) and No (count).
    The columns are Evaluating with a range 0 to 30, Reviewing with a range 0 to 30, and Agreement with system with a range 0 to 50.
    Below the bar charts are three point plots with range 1 to 5, aligned horizontally with the same columns as the bar charts.
    The single row is labeled Confidence (1-5).
  }%
  \label{fig:interactions}
\end{figure*}
}{
}

\medskip
\noindent\textbf{Interactions and answers. }
We used participants' interaction logs to compare usage in terms of turns taken, goals addressed, and answers when asked to evaluate and review their goals (Fig.~\ref{fig:interactions}).

Turns taken was similar between interfaces, between $3-5$ on average.
We observed\add{ no evidence for} OnGoal users having less goals confirmed\add{ (B: $3.8$ $[3.4, 4.2]$, O: $3.7$ $[3.3, 4.1]$)} and contradicted\add{ (B: $0.9$ $[0.6, 1.1]$, O: $0.7$ $[0.5, 0.9]$)}, while we observed\add{ weak evidence of} more goals being ignored\add{ (B: $1.3$ $[1, 1.7]$, O: $1.7$ $[1.3, 2.1]$)} than baseline users.
Further, the variability in goals being addressed, or the number of times a goal would switch between confirmed, contradicted, and ignored, was much wider (measured as the width of the $95$\% CI) when using OnGoal\add{ (B: $[5, 8.4]$, O: $[4.4, 12.2]$)}.
These results might point to OnGoal participants experimenting more with communication strategies to get the LLM to change its evaluation, with mixed results.

When asked to evaluate and review their conversational goals after each turn taken, OnGoal users felt the LLM addressed their goals more often and felt more confident in their answers than baseline users for both evaluating\add{ (B: $4.1$ $[3.8, 4.4]$, O: $4.5$ $[4.3, 4.8]$, strong evidence)} and reviewing\add{ (B: $4.3$ $[4.1, 4.6]$, O: $4.5$ $[4.3, 4.8]$, weak evidence)}.
Further, we looked at how often users agreed with the system evaluation.
\add{While both interfaces evaluated goals after each turn, evaluations were only visible to participants using OnGoal.}
We found participants tended to agree more with the system when using OnGoal, and their confidence was aligned with their answer --- when participants agreed with the system they were more confident\add{ (B: $4.2$ $[3.9, 4.5]$, O: $4.7$ $[4.6, 4.9]$, strong evidence)}, and conversely less confident if they disagreed\add{ (B: $4.2$ $[3.9, 4.5]$, O: $3.5$ $[2.4, 4.6]$, weak evidence)}.
This is likely due to the influence of seeing the evaluation on the screen.

\aptLtoX[graphic=no, type=html]{
\begin{figure}[!t]
\centering
\includegraphics[width=\linewidth]{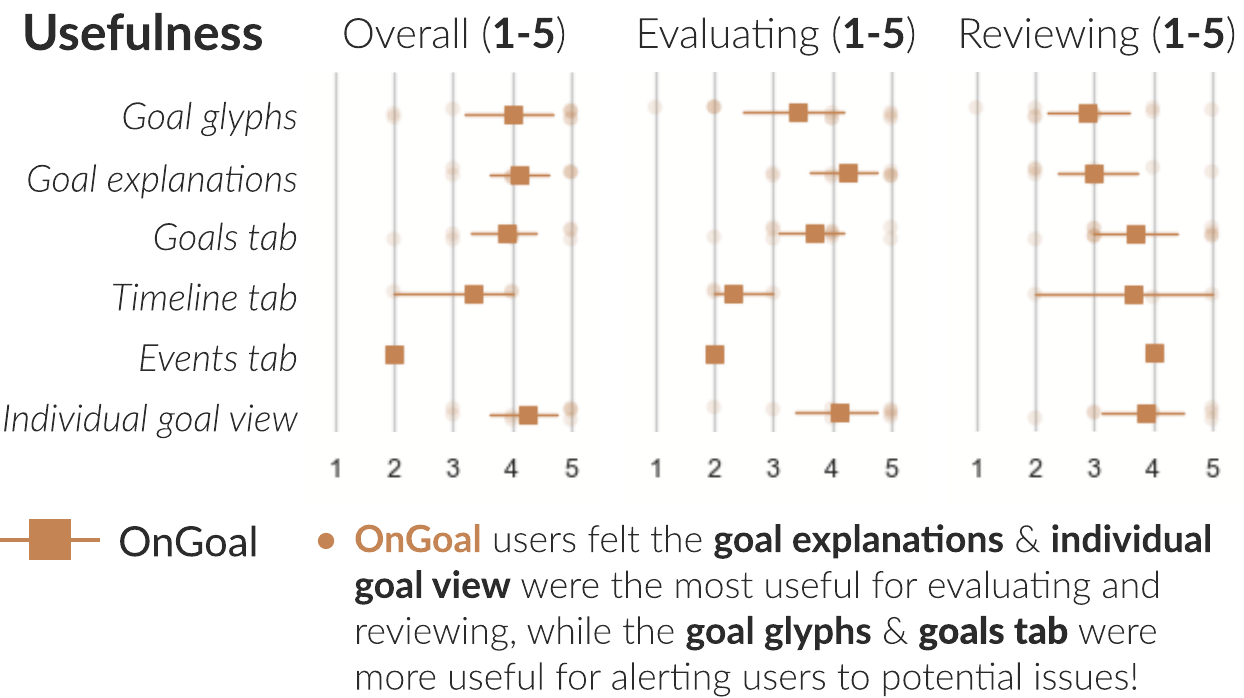}
  \caption{%
    Usefulness ratings for various OnGoal features.\strike{ The goal explanations and the individual goal view helped users to identify problematic LLM behaviors, while the goal glyphs and the tabs in the progress panel helped alert users to changes in their goal progress.}
  }%
  \Description{%
    Three side-by-side sub-figures of horizontal point plots with 95\% confidence intervals, each with a range 1 to 5.
    All point plots share six rows labeled Goal glyphs, Goal explanations, Goals tab, Timeline tab, Events tab, and Individual goal view, as well as one sub-row for OnGoal for each main row.
    The left sub-figure is labeled Overall (1-5), the middle sub-figure is labeled Evaluating (1-5), and the right sub-figure is labeled Reviewing (1-5).
  }%
  \label{fig:usefulness}
\end{figure}
}{
}

\medskip
\noindent\textbf{OnGoal features. }
We examined how participants used the features of OnGoal based on post-study survey feedback on usefulness (Fig.~\ref{fig:usefulness}) and think-aloud feedback.
 
Goal glyphs and the goals tab were self-reported more useful overall\add{ (goal glyphs: $4$ $[3.2, 4.7]$, goals tab: $3.9$ $[3.3, 4.4]$)} for alerting the user to potential issues over time such as a contradicted goal, and compared with evaluating\add{ (goal glyphs: $3.4$ $[2.5, 4.2]$, goals tab: $3.7$ $[3.1, 4.2]$, weak evidence)} or reviewing\add{ (goal glyphs: $2.9$ $[2.2, 3.6]$, goals tab: $3.7$ $[3.0, 4.4]$, weak evidence)}.
P$20$ gave an example for their reasoning: \textit{``It was more surprising to see red. That made me want to interact with the glyphs more. I could quickly find contradicting goals to fix them right away.''}

For evaluation, users predominantly found the goal explanations more useful than other features for evaluating\add{ ($4.2$ $[3.6, 4.8]$)}, helping them identify issues with how the LLM interpreted the goal in context of the response.
P$14$ would \textit{``click on the goal and immediately see why the goal was being satisfied and where in the response.''}
Depending on how difficult those issues were to resolve, users would then get details on demand in the individual goal view.
They considered this view more useful than other features for reviewing\add{ ($3.8$ $[3.1, 4.5]$)} the history of messages.
\add{The text highlighting techniques helped users identify key phrases and mismatching sentences, which served as signals for a lack of expected performance from the LLM.}
For P$19$, \textit{``the unique sentences and keyphrases helped me jump to different parts of the text to find evidence that align with the goal that I had.''}

The timeline and events tabs were the least useful features overall\add{ (timeline: $3$ $[2, 4]$, events: $2$ $[1.9, 2.1]$)} for this task.
Those participants that used them explained that they were not as useful when reviewing only a few fixed goals, and that they preferred these tabs more when reviewing the history of their own goals.

\medskip
\noindent\textbf{\add{OnGoal pipeline accuracy. }}
\add{Finally, we asked participants in the post-study survey to rate the accuracy of the different stages of the goal pipeline (infer, merge, evaluate). Because baseline users only had the infer stage active, they only rated that stage.}

\add{We found that both baseline and OnGoal users generally agreed on the performance of the infer stage (B: $3.6$ $[3.0, 4.2]$, O: $4.1$ $[3.6, 4.6]$, weak evidence). This provides confidence in our study setup that our conditions are comparable. OnGoal users self-reported higher accuracy for the infer and merge ($4$ $[3.0, 5.0]$) stages than for the evaluate stage ($2.9$ $[2.4, 3.4]$).
In particular, the difference between infer and evaluate was statistically significant. This may corroborate observations of emergent issues in our think-aloud feedback (Sect.~\ref{sec:emergent_issues}) --- some participants felt confused or misled by evaluations when they failed to align with their mental model of their own writing. This feedback from user-reported accuracy is important, as it reveals critical areas for improving goal evaluation, which we discuss in Sect.~\ref{sec:design_implications}.}

\aptLtoX[graphic=no, type=html]{
}{
\begin{table*}[!t]
    \caption{%
      Themes synthesized from think-aloud feedback.
    }%
    \Description{%
      Table with three columns and four rows.
      The first row is a header row.
      The first column in the header row is labeled "Theme", the second is labeled "Baseline", and the third is labeled "OnGoal".
      The last three rows of the first column are labels for the rows to the right; from top to bottom, Communicating goals, Maintaining awareness, and Identifying LLM issues.
      The last three rows of the second and third column are the data.
    }%
    \label{tab:thematic_analysis}
    \begin{tabular}{rp{0.38\linewidth}p{0.38\linewidth}}
        \toprule
            Theme & Baseline & OnGoal \\
        \midrule
            \multicolumn{1}{r}{Communicating goals} &
            Users often relied on and repeated the initial prompt to communicate goals, sometimes making little progress. &
            Users leveraged explanations to refine their goal communication and adapt their prompting strategies. \\
            \multicolumn{1}{r}{Maintaining awareness} &
            Users had to skim lengthy chat logs \& responses, decreasing confidence and increasing cognitive load. &
            Users offloaded cognition to goal glyphs and progress panel, strategically shifting their mental burden. \\
            \multicolumn{1}{r}{Identifying LLM issues} &
            Users manually compared messages for indicators of goal progress to learn why the LLM failed. &
            Users leveraged highlighting and progress panel to detect, understand, and monitor goal alignment. \\
        \bottomrule
    \end{tabular}
\end{table*}
}

\aptLtoX[graphic=no, type=html]{
\begin{table*}[!t]
    \caption{%
      Themes synthesized from think-aloud feedback.
    }%
    \Description{%
      Table with three columns and four rows.
      The first row is a header row.
      The first column in the header row is labeled "Theme", the second is labeled "Baseline", and the third is labeled "OnGoal".
      The last three rows of the first column are labels for the rows to the right; from top to bottom, Communicating goals, Maintaining awareness, and Identifying LLM issues.
      The last three rows of the second and third column are the data.
    }%
    \label{tab:thematic_analysis}
    \begin{tabular}{rp{0.38\linewidth}p{0.38\linewidth}}
        \toprule
            Theme & Baseline & OnGoal \\
        \midrule
            \multicolumn{1}{r}{Communicating goals} &
            Users often relied on and repeated the initial prompt to communicate goals, sometimes making little progress. &
            Users leveraged explanations to refine their goal communication and adapt their prompting strategies. \\
            \multicolumn{1}{r}{Maintaining awareness} &
            Users had to skim lengthy chat logs \& responses, decreasing confidence and increasing cognitive load. &
            Users offloaded cognition to goal glyphs and progress panel, strategically shifting their mental burden. \\
            \multicolumn{1}{r}{Identifying LLM issues} &
            Users manually compared messages for indicators of goal progress to learn why the LLM failed. &
            Users leveraged highlighting and progress panel to detect, understand, and monitor goal alignment. \\
        \bottomrule
    \end{tabular}
\end{table*}
}{
}

\subsubsection{Thematic analysis}
\label{sec:thematic_analysis}
We then organized participants' think-aloud feedback thematically based on emergent issues they encountered when interacting with the LLM over multiple conversational turns, and corroborated feedback with evidence from the usability insights in Sect.~\ref{sec:usability_findings}.
We further deductively aligned our themes of participant feedback with our design challenges (\textbf{C}) in Sect.~\ref{sec:methodology}.

Baseline users shared several challenges including miscommunication of goals, excessive effort spent reading the chat, and uncertainty about how their goals were being addressed.
In contrast, OnGoal users employed more diverse strategies to overcome miscommunication, spent more time and effort on evaluating and reviewing goals, and identified more ways to make consistent progress, fostering greater confidence.

\medskip
\noindent\textbf{Communicating goals. }
Participants in both conditions struggled with getting the LLM to understand their goals, as anticipated from the design challenge \textbf{C1}.
However, baseline users particularly struggled with \textit{how} and \textit{when} to effectively communicate their goals more than OnGoal users.
\add{We expected baseline users to spend more time evaluating and reviewing (Sect.~\ref{sec:usability_findings}) with less information to process, giving them more time to experiment with \textit{how} and \textit{when} to communicate goals. Instead, OnGoal's simple feedback loop more often led users to adapt goal communication, even when lacking suggestions on what to fix. Designing interfaces to encourage more active, dynamic dialogue may help users overcome goal misalignment in similar ways to improving modeling accuracy.}

For example, we observed baseline users often resorting to long and comprehensive initial prompts in an attempt to cover all their goals.
However, this approach frequently led the LLM to misinterpret or ignore goals, as P$4$ lamented: \textit{``It was often not congruent with the prompt at the end of the response.''}
In contrast, OnGoal's in-situ explanations encouraged more iterative goal refinement.
By explicitly flagging which goals were misinterpreted or overlooked and where, the system encouraged users to adjust prompts iteratively, directing the LLM's attention to specific sections of the response.
For example, this helped P$18$ become unstuck on how to keep moving forward: \textit{``The interface outlines where the agreement or disagreement occurs... I could then compare my own mental model to the system's model to help me move on.''}

Moreover, when baseline users felt the LLM was ``stuck'' or drifting from their intent, they often repeated similar prompts with little success, leading to frustration and stagnation.
P$2$ tried this strategy and found that \textit{``the LLM response's problems have been pretty stubborn... I need multiple [turns] to drift [the LLM] away from these issues.''}
This stagnation is likely reflected in the lower variability of goal status changes in Fig.~\ref{fig:interactions}.
With OnGoal, users instead utilized explanation cues to target specific parts of the output or rephrase requests based on what the system flagged as problematic.
For example, P$13$ targeted specific sections of the response in follow-up messages: \textit{``I want the LLM to align the goal with sections of the article that naturally would have more rich imagery.''}

Additionally, some baseline participants repeatedly attempted to force all goals to be perfectly aligned in every response, even when the LLM clearly failed to understand or adapt.
This might explain the higher rate of ``no'' answers when evaluating and reviewing goals compared with OnGoal in Fig.~\ref{fig:interactions}.
OnGoal flipped this dynamic: explanations often encouraged users to reconsider or revise their goals based on the LLM's interpretation.
Some, like P$12$, reported gaining more trust in the LLM's judgment after reviewing its reasoning, leading to more flexible and productive interactions: \textit{``It's surprising that the LLM recognizes the conflicts in the request, and generates different versions correspondingly. This is clever.''}
We also observed OnGoal's explanations help users better understand the LLM's intent and reassess their own personal judgments.
Specifically, some onGoal participants changed their evaluations from ``no'' to ``yes'' after seeing evidence they had initially missed, showing how explanation-driven feedback fostered a deeper, shared understanding of task progress.
For example, P$18$ changed their initial ``no'' evaluation based on an LLM explanation: \textit{``I actually agree with the system, the message does have examples that changes my mind. I do think the goal is satisfied.''}
Such instances likely contributed to the higher proportion of ``yes'' answers we observed in Fig.~\ref{fig:interactions}.

\medskip
\noindent\textbf{Maintaining awareness. }
Users in both interface conditions encountered sensemaking challenges, such as increased effort needed to parse lengthy LLM responses and chat logs, as posited by \textbf{C2}.
However, OnGoal's text highlighting provided cues that helped users know \textit{where} to shift their focus, saving them time and effort in the long run over baseline users (Fig.~\ref{fig:performance}).

With the baseline interface, users described that scanning long responses and chat logs was time-consuming and mentally taxing.
The LLM also made mistakes that were difficult to parse, as P$7$ points out: \textit{``I spent a lot of time reading the text conversation. I couldn't pay attention to quality... I can't believe I didn't even notice the grammatical errors until the end.''}
P$10$ manually scanned for goal-related phrases and indicators, recreating functionality that OnGoal provided directly.
The manual effort likely corroborates the lower confidence self-reported in Fig.~\ref{fig:performance}, especially during review, as P$6$ explains: \textit{``It's a bit too long to go through all the responses that I have previously, so I'm decreasing my confidence score.''}

In contrast, OnGoal shifted the burden away from exhaustive reading by visualizing signals of goal progress directly.
The goal glyphs and progress panel were predominantly used to alert users to inconsistencies between their prompt and the goals.
P$17$ used this feedback to reduce their cognitive load: \textit{``I didn't have to actively and consistently think about the goals all the time... [OnGoal] reduced the cognitive load for me as a writer.''}
OnGoal's features seemed to empower users to use their mental energy more strategically.
P$11$ used the the goal glyphs and timeline tab to offload cognition to the system, only needing to check in sporadically when issues arose: \textit{``I can just check the response for which goals I think it should be satisfying.''}
P$13$ used the goal-tracking features as gauges for when to turn up or down a particular goal: \textit{``Now that I've seen the response, I want to focus on "turning the knobs up and down" on different goals.''}
The popularity of these OnGoal features also aligns with the usefulness feedback and ratings reported in Fig.~\ref{fig:usefulness}.

\medskip
\noindent\textbf{Identifying LLM issues. }
While the LLM would sometimes forget goals or ignore requests in both conditions per \textbf{C3}, OnGoal could visualize textual patterns related to addressing goals across messages, helping users better understand \textit{why} the LLM was underperforming and avoid derailing the entire conversation.

Specifically, baseline users often struggled to assess goal consistency over time --- a potential reason for more ``no'' responses when reviewing goals (Fig.~\ref{fig:interactions}).
For example, P$5$ had to scan what GPT changed between messages to determine progress: \textit{``The GPT didn't really change a lot of its responses depending on my prompting... if there is always a similar response, it means the response is consistent, but on the other hand, the GPT is not polishing the answer according to my prompt.''}
In contrast, OnGoal's inline highlighting and sentence comparison features helped users discern between helpful consistency and unproductive repetition.
Several strategies emerged: P$18$ looked for key phrases in highlighted text to confirm their requests had been addressed, P$12$ checked if unique sentences changed or stayed stagnant over time, and P$20$ compared both new and repeated sentences to track progress: \textit{``During goal review, I pulled out the unique sentences to see if it was generating new things, and then similar sentences to see if it kept what I liked.''}

While reviewing multiple goals remained cognitively demanding in both conditions, users in the baseline condition reported lower confidence and spent less time engaging with the task (Fig.~\ref{fig:performance}).
A lack of transparency features for LLM behaviors may be the cause, as many baseline participants spent time scrolling through lengthy chat logs to understand what went wrong.
To save time and effort, some like P$4$ implicitly trusted the LLM to remember their goals and conversational history.
Others like P$1$ and P$8$ explicitly requested transparency features that we integrated into OnGoal -- LLM goal evaluations and text highlighting between messages.
OnGoal's text highlighting and progress panel vitally reduced the burden of reviewing goal progress.
For example, P$14$ glanced at the timeline to see if anything had become contradicted (i.e., red): \textit{``Once I determined my feeling about the system performance, I relied more on glancing at [the timeline tab] to help with reviewing.''}
P$11$ uniquely asked the LLM to summarize goal progress at the end of the task, then used the text highlighting features to identify where it had diverged from specific goals and why.
These features likely also helped OnGoal users save time reading (as seen in Fig.~\ref{fig:performance}).

\subsubsection{Emergent issues}
\label{sec:emergent_issues}
Despite the benefits of goal tracking and visualization we observed, several new issues emerged uniquely from the OnGoal condition.
Some participants felt that the system's goal explanations could be contentious.
For instance, P$15$ felt the goal explanations encouraged them to see their writing as more objective: \textit{``There is no right or wrong in terms of what I want for my plan. But, there is now a right and wrong in terms of goals for the writing... I was fighting the LLM for what I wanted.''}
Other participants were confused when the LLM gave examples that seemed to contradict its own evaluation.
This led P$14$ to increase their mental effort: \textit{``There were times I had to go against the system, especially when the system was disagreeing with me, and I was spending a lot of effort.''}
Some felt the LLM evaluations were not always clear, and could lead to confusion about how the LLM was evaluating the response, including P$16$ (\textit{``The glyph in the message cited a reason for ignoring the goal, but it wasn't obvious why it came to that assessment.''}) and P$12$ (\textit{``Some evidence were considered as satisfying the goal, some of the same are considered ignoring the goal.''}).
This usually came down to a mismatch in expectations, which P$16$ explained: \textit{``When the goal said “ignore” I was confused, because the part that was relevant to me was still satisfied.''}
These observations highlight that while the transparency features provided by OnGoal can empower users, they also introduce interpretive complexity, which may require extra justification effort.

\section{Discussion}
\label{sec:discussion}

Our work contributes to a growing area of research which explores the cognitive gap between end users and AI in task-driven settings \cite{subramonyam2023bridging, zamfirescu2023johnny, tankelevitch2024metacognitive}.
Specifically, users of AI agents like LLMs can struggle to express their wants to the AI and align their expectations with how the AI interprets their goals, making it difficult to verify whether AI-generated content meets their objectives \cite{Lee:2025:GenAICriticalThinking}.
To address this gap, recent works have explored how improved UI designs and workflows can assist users in leveraging AI for writing \cite{laban2024beyond}, image editing \cite{masson2024directgpt}, and sensemaking \cite{Suh:2023:Sensecape}.
Our work specifically explores the benefits of improving a UI by explicitly encoding and showing users' goals, towards addressing the general limitations of LLMs in conversational contexts.
In Sect.~\ref{sec:design_implications}, we synthesize our study findings on goal tracking and visualization into design implications for future interfaces that foster increased human-AI collaboration and creativity while keeping users engaged and empowered.
Then, in Sect.~\ref{sec:limitations_future}, we discuss limitations of our evaluation and future work studying interaction with LLMs in multi-turn dialogue.

\subsection{Implications For Design}
\label{sec:design_implications}
We developed design implications that build off of participants' think-aloud feedback and our own observations of how interacting with LLMs in multi-turn conversation can be improved.

\medskip
\noindent\textbf{Enable multiple ways to communicate goals.  }
Formalizing conversational goals in multi-turn dialogue can give users greater control over their interactions with LLMs.
While baseline users typically listed all goals in the initial prompt, often with mixed results, OnGoal users employed diverse strategies for \textit{how} and \textit{when} to communicate goals, helping them navigate miscommunication more effectively.
For example, participants varied \textit{how} they communicated goals with the LLM: (1) some put all goals in the initial prompt up front, similar to the baseline condition; (2) some put a few goals in a time, such as conflicting pairs; (3) and some never put goals in, only asking the LLM to adjust its responses on a case-by-case basis.
We also observed variations in \textit{when} goals were discussed with the LLM: (1) \textit{proactive}, where participants explicitly asked for the LLM to address target goals; and (2) \textit{reactive}, where participants allowed the LLM to fulfill the task implicitly without having specific goals, and used the system features to identify if changes were needed through specific instructions. 
These strategies highlight a need for personalized workflows that align users' preferences for inferring and merging goals across task stages \cite{wang2024task}.
Additionally, users' goal-setting priorities shifted across different stages of the task, reflecting a shift in users' mental models \cite{mahmood2023llm}.
For example, when goals were unspecified, users often prioritized narrative flow, while specific goals led users to concentrate on more targeted wording edits.
Building off of OnGoal's on-demand goal inference and evaluation, future systems could adapt dynamically to evaluate goals where users are focusing on and support flexible goal-setting throughout the dialogue.

\medskip
\noindent\textbf{Visualize where goal evaluations align with users' focus.  }
In supporting multiple ways for users to communicate their goals, visualizations play a key role in revealing how well responses align with goals over time.
We observed that baseline users often relied on their initial prompt structure to scaffold their process for evaluating and reviewing progress on goals.
For example, in follow-up prompts, users would only check the LLM response against what they had originally asked for, overlooking missing goals or contradictions unless those were explicitly stated in the initial prompt.
In contrast, OnGoal users leveraged the examples and text highlighting to identify where and how goals were not being addressed by the LLM, resulting in more informed assessment and better goal alignment over time.
These findings may point at design opportunities for visualizations to provide a layer of understanding, helping users monitor the evolving relationship between their goals and LLM responses.
For example, some OnGoal users requested summary visualizations of evolving themes and key ideas across message blocks, like in ThemeRiver \cite{Havre:2000:ThemeRiver}.
To indicate interest, direct manipulation could be leveraged to highlight specific clauses in a response \cite{masson2024directgpt}, and then trace that highlight across messages to see if similar sentences or themes are addressed by the LLM or ignored.
Visualizations should further highlight the user's interests and supply more in-depth explanations using LLMs that can reason about why a goal was evaluated the way it was.
At the same time, care should be taken to carefully annotate the purpose of the visualizations, as OnGoal users would sometimes interpret highlighting as a proxy for determining where the LLM was focusing its attention.

\medskip
\noindent\textbf{Design goal alerts and snapshots to further offload cognition.  }
Compared with the baseline, we saw OnGoal users report higher levels of confidence when evaluating and reviewing for similar or less effort.
This may indicate that conversational interaction with LLMs is enhanced by having the system offload tasks from the user such as tracking conversational goals over time.
For example, participants reported that the goal glyphs were useful for altering them when issues were present such as contradictions, freeing them up to think about how to make progress on their other goals.
To help users rely less on their memory and keep them in the loop longer, goal alerts could be configurable, such as alerts when certain combinations of goals conflict or too many have been merged.
While we provide summary visualizations in the timeline and events views, they lacked the ability to trace individual text phrases such as key words or themes throughout.
To facilitate this, a summary snapshot could report progress in terms of salient themes addressed, or key phrases, where the system synthesizes the progress on completing goals automatically and lists what is left for the user to accomplish (i.e., ``did I get everything?'').
This is particularly important based on feedback we got that not all messages in a conversation were useful for all goals.
By summarizing progress, users could see potentially conflicting messages without needing to skim the conversation and take focus away from working on goals.

\medskip
\noindent\textbf{Support user feedback on goal evaluation.}
Goal evaluation is often subjective, varying according to user's individual interpretations, preferences, and needs.
We observed that users occasionally disagreed with LLM's assessments and experienced tension when they were unable to influence how their goals were interpreted or judged.
It was not always clear to users how the LLM arrived at an evaluation, and further, users could not give the system feedback to improve or personalize these judgments.
Future work should explore external human-in-the-loop control and feedback for personalizing the goal pipeline, such as giving evaluations a ``thumbs up'' or ``thumbs down'' to adapt the evaluation style towards users' preferences over time \cite{li2024chathf, shi2024wildfeedback}.
More interactive methods could be used, such as combining LLM-as-a-judge \cite{Zheng:2023:ChatbotArena} with expanded visualizations to provide examples of users' intention for updating the LLM evaluation prompts, like in Scattershot \cite{Wu:2023:ScatterShot}.
Evaluations could further be regenerated or edited by the user to better keep track of whether the system addressed goals when reviewing the conversation later on.

\subsection{Limitations and Future Work}
\label{sec:limitations_future}
\add{Our system implementation has several limitations that reveal promising future research directions.}
\add{To better understand how OnGoal's goal pipeline affected human-AI alignment in LLM chat interfaces, we collected user-reported accuracy, which helped contextualize important findings on user experience, task outcomes and behavioral insights.}\strike{ This work focused on investigating our proof-of-concept UI for tracking and visualizing conversational goals, rather than aiming to evaluate the quality of the goal pipeline.}
\add{However, quantitatively evaluating our pipeline's accuracy, such as on expert-annotated benchmarks, remains untested. Complimenting user-reported accuracy with expert-annotated precision could provide further insights into the effects of goal tracking on  human-AI alignment.}
For example, how the \add{precision}\strike{ quality} of the goal pipeline might affect a user's ability to evaluate and review goals remains unclear.
\add{In addition to benchmarks,} future work should also investigate how prompt design and alternative, potentially more robust, LLM architectures could improve goal tracking.
At the same time, including smaller or less capable models such as compressed or distilled LLMs would help assess the generalizability of our findings.
\add{Finally, it is unclear how support for other goal types might affect user behaviors. In cases where global tracking alone may be insufficient for complex multi‑section artifacts, fine-grained and local goals could help bridge usability gaps.}

In our evaluation, learning effects with OnGoal were inconclusive.
For example, some participants reported that reviewing became easier over time as they grew more familiar with the LLM, regardless of interface features.
It remains unclear how much OnGoal contributed beyond this growing familiarity.
Further, our evaluation on a writing task and a baseline chat interface was useful to highlight interface-related insights.
However, it is unclear whether additional cognitive and sensemaking effects might emerge in other tasks like copywriting, data analysis, personal assistance, programming, or learning.
In exploring other tasks, our approach could also extend to human-to-human conversation, such as in online interactions in virtual meeting spaces where LLMs could assist in goal tracking and visualization.
Lastly, we did not examine whether our system encourages users to create more or more diverse goals.
Some feedback suggested the timeline and events views may be especially helpful in open-ended contexts.
Future work should explore dynamic goal tracking over time in longitudinal studies with open-ended goal setting.

\section{Conclusion}
\label{sec:conclusion}

Using LLMs as conversational agents is opening the door to new opportunities for supporting users in solving complex tasks.
As these systems are deployed in diverse environments with unknown LLM capabilities, interfaces can provide the critical infrastructure needed to facilitate stronger communication with LLM agents.
Our work contributes study results and design implications for conversational goal tracking and visualization as an avenue for enabling more efficient and resilient LLM chat interface experiences.
By developing OnGoal, we uncovered specific insights into how users of chat interfaces can handle miscommunication with LLMs, make sense of long and complex chat histories, and identify problematic LLM behaviors using interactive visualizations.
Future designers and engineers should explore more ways to support personalized experiences in LLM chat interfaces, including new methods for communicating goals and feedback-driven goal evaluations.

\begin{acks}
  This research was conducted during an internship at Adobe Research.
  We thank the Adobe Research EEL group for their guidance and support throughout the work.
\end{acks}

\bibliographystyle{ACM-Reference-Format}
\bibliography{main}


\begin{thebibliography}{56}


\ifx \showCODEN    \undefined \def \showCODEN     #1{\unskip}     \fi
\ifx \showISBNx    \undefined \def \showISBNx     #1{\unskip}     \fi
\ifx \showISBNxiii \undefined \def \showISBNxiii  #1{\unskip}     \fi
\ifx \showISSN     \undefined \def \showISSN      #1{\unskip}     \fi
\ifx \showLCCN     \undefined \def \showLCCN      #1{\unskip}     \fi
\ifx \shownote     \undefined \def \shownote      #1{#1}          \fi
\ifx \showarticletitle \undefined \def \showarticletitle #1{#1}   \fi
\ifx \showURL      \undefined \def \showURL       {\relax}        \fi
\providecommand\bibfield[2]{#2}
\providecommand\bibinfo[2]{#2}
\providecommand\natexlab[1]{#1}
\providecommand\showeprint[2][]{arXiv:#2}

\bibitem[Arawjo et~al\mbox{.}(2024)]%
        {Arawjo:2024:ChainForge}
\bibfield{author}{\bibinfo{person}{Ian Arawjo}, \bibinfo{person}{Chelse
  Swoopes}, \bibinfo{person}{Priyan Vaithilingam}, \bibinfo{person}{Martin
  Wattenberg}, {and} \bibinfo{person}{Elena~L. Glassman}.}
  \bibinfo{year}{2024}\natexlab{}.
\newblock \showarticletitle{ChainForge: A Visual Toolkit for Prompt Engineering
  and LLM Hypothesis Testing}. In \bibinfo{booktitle}{\emph{Proceedings of the
  2024 CHI Conference on Human Factors in Computing Systems}} (Honolulu, HI,
  USA) \emph{(\bibinfo{series}{CHI '24})}. \bibinfo{publisher}{Association for
  Computing Machinery}, \bibinfo{address}{New York, NY, USA}, Article
  \bibinfo{articleno}{304}, \bibinfo{numpages}{18}~pages.
\newblock
\showISBNx{9798400703300}
\href{https://doi.org/10.1145/3613904.3642016}{doi:\nolinkurl{10.1145/3613904.3642016}}


\bibitem[Ashby et~al\mbox{.}(2024)]%
        {Ashby:2024:LLMConversationTopic}
\bibfield{author}{\bibinfo{person}{Trevor Ashby}, \bibinfo{person}{Adithya
  Kulkarni}, \bibinfo{person}{Jingyuan Qi}, \bibinfo{person}{Minqian Liu},
  \bibinfo{person}{Eunah Cho}, \bibinfo{person}{Vaibhav Kumar}, {and}
  \bibinfo{person}{Lifu Huang}.} \bibinfo{year}{2024}\natexlab{}.
\newblock \showarticletitle{Towards Effective Long Conversation Generation with
  Dynamic Topic Tracking and Recommendation}. In
  \bibinfo{booktitle}{\emph{Proceedings of the 17th International Natural
  Language Generation Conference}}, \bibfield{editor}{\bibinfo{person}{Saad
  Mahamood}, \bibinfo{person}{Nguyen~Le Minh}, {and} \bibinfo{person}{Daphne
  Ippolito}} (Eds.). \bibinfo{publisher}{Association for Computational
  Linguistics}, \bibinfo{address}{Tokyo, Japan}, \bibinfo{pages}{540--556}.
\newblock
\urldef\tempurl%
\url{https://aclanthology.org/2024.inlg-main.43/}
\showURL{%
\tempurl}


\bibitem[Baker et~al\mbox{.}(2009)]%
        {Baker:2009:VisEnhancesSensemaking}
\bibfield{author}{\bibinfo{person}{Jeff Baker}, \bibinfo{person}{Donald Jones},
  {and} \bibinfo{person}{Jim Burkman}.} \bibinfo{year}{2009}\natexlab{}.
\newblock \showarticletitle{Using visual representations of data to enhance
  sensemaking in data exploration tasks}.
\newblock \bibinfo{journal}{\emph{Journal of the Association for Information
  Systems}} \bibinfo{volume}{10}, \bibinfo{number}{7} (\bibinfo{year}{2009}),
  \bibinfo{pages}{2}.
\newblock


\bibitem[Bjork et~al\mbox{.}(2011)]%
        {bjork2011making}
\bibfield{author}{\bibinfo{person}{Elizabeth~L Bjork},
  \bibinfo{person}{Robert~A Bjork}, {et~al\mbox{.}}}
  \bibinfo{year}{2011}\natexlab{}.
\newblock \showarticletitle{Making things hard on yourself, but in a good way:
  Creating desirable difficulties to enhance learning}.
\newblock \bibinfo{journal}{\emph{Psychology and the real world: Essays
  illustrating fundamental contributions to society}} \bibinfo{volume}{2},
  \bibinfo{number}{59-68} (\bibinfo{year}{2011}).
\newblock


\bibitem[Bostock et~al\mbox{.}(2011)]%
        {Bostock:2011:D3}
\bibfield{author}{\bibinfo{person}{Michael Bostock}, \bibinfo{person}{Vadim
  Ogievetsky}, {and} \bibinfo{person}{Jeffrey Heer}.}
  \bibinfo{year}{2011}\natexlab{}.
\newblock \showarticletitle{D$^3$ Data-Driven Documents}.
\newblock \bibinfo{journal}{\emph{IEEE Transactions on Visualization and
  Computer Graphics}} \bibinfo{volume}{17}, \bibinfo{number}{12}
  (\bibinfo{year}{2011}), \bibinfo{pages}{2301--2309}.
\newblock
\href{https://doi.org/10.1109/TVCG.2011.185}{doi:\nolinkurl{10.1109/TVCG.2011.185}}


\bibitem[Boyatzis(1998)]%
        {Boyatzis:1998:ThematicAnalysis}
\bibfield{author}{\bibinfo{person}{R.E. Boyatzis}.}
  \bibinfo{year}{1998}\natexlab{}.
\newblock \bibinfo{booktitle}{\emph{Transforming Qualitative Information:
  Thematic Analysis and Code Development}}.
\newblock \bibinfo{publisher}{SAGE Publications}.
\newblock
\showISBNx{9780761909613}
\showLCCN{97045405}


\bibitem[Bursztyn et~al\mbox{.}(2021)]%
        {Bursztyn:2021:LLMConvoRec}
\bibfield{author}{\bibinfo{person}{Victor~S. Bursztyn},
  \bibinfo{person}{Jennifer Healey}, \bibinfo{person}{Eunyee Koh},
  \bibinfo{person}{Nedim Lipka}, {and} \bibinfo{person}{Larry Birnbaum}.}
  \bibinfo{year}{2021}\natexlab{}.
\newblock \showarticletitle{Developing a Conversational Recommendation
  Systemfor Navigating Limited Options}. In \bibinfo{booktitle}{\emph{Extended
  Abstracts of the 2021 CHI Conference on Human Factors in Computing Systems}}
  (Yokohama, Japan) \emph{(\bibinfo{series}{CHI EA '21})}.
  \bibinfo{publisher}{Association for Computing Machinery},
  \bibinfo{address}{New York, NY, USA}, Article \bibinfo{articleno}{309},
  \bibinfo{numpages}{6}~pages.
\newblock
\showISBNx{9781450380959}
\href{https://doi.org/10.1145/3411763.3451596}{doi:\nolinkurl{10.1145/3411763.3451596}}


\bibitem[Chen et~al\mbox{.}(2025)]%
        {Chen:2025:StuGPTViz}
\bibfield{author}{\bibinfo{person}{Zixin Chen}, \bibinfo{person}{Jiachen Wang},
  \bibinfo{person}{Meng Xia}, \bibinfo{person}{Kento Shigyo},
  \bibinfo{person}{Dingdong Liu}, \bibinfo{person}{Rong Zhang}, {and}
  \bibinfo{person}{Huamin Qu}.} \bibinfo{year}{2025}\natexlab{}.
\newblock \showarticletitle{StuGPTViz: A Visual Analytics Approach to
  Understand Student-ChatGPT Interactions}.
\newblock \bibinfo{journal}{\emph{IEEE Transactions on Visualization and
  Computer Graphics}} \bibinfo{volume}{31}, \bibinfo{number}{1}
  (\bibinfo{year}{2025}), \bibinfo{pages}{908--918}.
\newblock
\href{https://doi.org/10.1109/TVCG.2024.3456363}{doi:\nolinkurl{10.1109/TVCG.2024.3456363}}


\bibitem[Deriu et~al\mbox{.}(2021)]%
        {Deriu:2021:SurveyDialogueEvaluation}
\bibfield{author}{\bibinfo{person}{Jan Deriu}, \bibinfo{person}{Alvaro
  Rodrigo}, \bibinfo{person}{Arantxa Otegi}, \bibinfo{person}{Guillermo
  Echegoyen}, \bibinfo{person}{Sophie Rosset}, \bibinfo{person}{Eneko Agirre},
  {and} \bibinfo{person}{Mark Cieliebak}.} \bibinfo{year}{2021}\natexlab{}.
\newblock \showarticletitle{Survey on evaluation methods for dialogue systems}.
\newblock \bibinfo{journal}{\emph{Artificial Intelligence Review}}
  \bibinfo{volume}{54}, \bibinfo{number}{1} (\bibinfo{date}{01 Jan}
  \bibinfo{year}{2021}), \bibinfo{pages}{755--810}.
\newblock
\showISSN{1573-7462}
\href{https://doi.org/10.1007/s10462-020-09866-x}{doi:\nolinkurl{10.1007/s10462-020-09866-x}}


\bibitem[Dragicevic(2016)]%
        {Dragicevic:2016:HCIStats}
\bibfield{author}{\bibinfo{person}{Pierre Dragicevic}.}
  \bibinfo{year}{2016}\natexlab{}.
\newblock \bibinfo{booktitle}{\emph{Fair Statistical Communication in HCI}}.
\newblock \bibinfo{publisher}{Springer International Publishing},
  \bibinfo{address}{Cham}, \bibinfo{pages}{291--330}.
\newblock
\showISBNx{978-3-319-26633-6}
\href{https://doi.org/10.1007/978-3-319-26633-6_13}{doi:\nolinkurl{10.1007/978-3-319-26633-6_13}}


\bibitem[Duan et~al\mbox{.}(2023)]%
        {duan2023botchat}
\bibfield{author}{\bibinfo{person}{Haodong Duan}, \bibinfo{person}{Jueqi Wei},
  \bibinfo{person}{Chonghua Wang}, \bibinfo{person}{Hongwei Liu},
  \bibinfo{person}{Yixiao Fang}, \bibinfo{person}{Songyang Zhang},
  \bibinfo{person}{Dahua Lin}, {and} \bibinfo{person}{Kai Chen}.}
  \bibinfo{year}{2023}\natexlab{}.
\newblock \showarticletitle{Botchat: Evaluating llms' capabilities of having
  multi-turn dialogues}.
\newblock \bibinfo{journal}{\emph{arXiv preprint arXiv:2310.13650}}
  (\bibinfo{year}{2023}).
\newblock


\bibitem[Ericsson and Simon(1984)]%
        {Ericsson:1984:ProtocolAnalysis}
\bibfield{author}{\bibinfo{person}{K~Anders Ericsson} {and}
  \bibinfo{person}{Herbert~A Simon}.} \bibinfo{year}{1984}\natexlab{}.
\newblock \bibinfo{booktitle}{\emph{Protocol analysis: Verbal reports as
  data.}}
\newblock \bibinfo{publisher}{the MIT Press}.
\newblock


\bibitem[Fu et~al\mbox{.}(2018)]%
        {Fu:2018:TCal}
\bibfield{author}{\bibinfo{person}{Siwei Fu}, \bibinfo{person}{Jian Zhao},
  \bibinfo{person}{Hao~Fei Cheng}, \bibinfo{person}{Haiyi Zhu}, {and}
  \bibinfo{person}{Jennifer Marlow}.} \bibinfo{year}{2018}\natexlab{}.
\newblock \showarticletitle{T-Cal: Understanding Team Conversational Data with
  Calendar-based Visualization}. In \bibinfo{booktitle}{\emph{Proceedings of
  the 2018 CHI Conference on Human Factors in Computing Systems}} (Montreal,
  QC, Canada) \emph{(\bibinfo{series}{CHI '18})}.
  \bibinfo{publisher}{Association for Computing Machinery},
  \bibinfo{address}{New York, NY, USA}, \bibinfo{pages}{1–13}.
\newblock
\showISBNx{9781450356206}
\href{https://doi.org/10.1145/3173574.3174074}{doi:\nolinkurl{10.1145/3173574.3174074}}


\bibitem[Gao et~al\mbox{.}(2024)]%
        {Gao:2024:HumanLLMInteractionModes}
\bibfield{author}{\bibinfo{person}{Jie Gao}, \bibinfo{person}{Simret~Araya
  Gebreegziabher}, \bibinfo{person}{Kenny Tsu~Wei Choo}, \bibinfo{person}{Toby
  Jia-Jun Li}, \bibinfo{person}{Simon~Tangi Perrault}, {and}
  \bibinfo{person}{Thomas~W Malone}.} \bibinfo{year}{2024}\natexlab{}.
\newblock \showarticletitle{A Taxonomy for Human-LLM Interaction Modes: An
  Initial Exploration}. In \bibinfo{booktitle}{\emph{Extended Abstracts of the
  2024 CHI Conference on Human Factors in Computing Systems}} (Honolulu, HI,
  USA) \emph{(\bibinfo{series}{CHI EA '24})}. \bibinfo{publisher}{Association
  for Computing Machinery}, \bibinfo{address}{New York, NY, USA}, Article
  \bibinfo{articleno}{24}, \bibinfo{numpages}{11}~pages.
\newblock
\showISBNx{9798400703317}
\href{https://doi.org/10.1145/3613905.3650786}{doi:\nolinkurl{10.1145/3613905.3650786}}


\bibitem[Gero et~al\mbox{.}(2024)]%
        {Gero:2024:SensemakingLLMsAtScale}
\bibfield{author}{\bibinfo{person}{Katy~Ilonka Gero}, \bibinfo{person}{Chelse
  Swoopes}, \bibinfo{person}{Ziwei Gu}, \bibinfo{person}{Jonathan~K.
  Kummerfeld}, {and} \bibinfo{person}{Elena~L. Glassman}.}
  \bibinfo{year}{2024}\natexlab{}.
\newblock \showarticletitle{Supporting Sensemaking of Large Language Model
  Outputs at Scale}. In \bibinfo{booktitle}{\emph{Proceedings of the 2024 CHI
  Conference on Human Factors in Computing Systems}} (Honolulu, HI, USA)
  \emph{(\bibinfo{series}{CHI '24})}. \bibinfo{publisher}{Association for
  Computing Machinery}, \bibinfo{address}{New York, NY, USA}, Article
  \bibinfo{articleno}{838}, \bibinfo{numpages}{21}~pages.
\newblock
\showISBNx{9798400703300}
\href{https://doi.org/10.1145/3613904.3642139}{doi:\nolinkurl{10.1145/3613904.3642139}}


\bibitem[Gupta et~al\mbox{.}(2024)]%
        {Gupta:2024:LLMTaskSwitching}
\bibfield{author}{\bibinfo{person}{Akash Gupta}, \bibinfo{person}{Ivaxi Sheth},
  \bibinfo{person}{Vyas Raina}, \bibinfo{person}{Mark Gales}, {and}
  \bibinfo{person}{Mario Fritz}.} \bibinfo{year}{2024}\natexlab{}.
\newblock \showarticletitle{{LLM} Task Interference: An Initial Study on the
  Impact of Task-Switch in Conversational History}. In
  \bibinfo{booktitle}{\emph{ICML 2024 Workshop on Foundation Models in the
  Wild}}.
\newblock
\urldef\tempurl%
\url{https://openreview.net/forum?id=WnMcWR9n3P}
\showURL{%
\tempurl}


\bibitem[Havre et~al\mbox{.}(2000)]%
        {Havre:2000:ThemeRiver}
\bibfield{author}{\bibinfo{person}{S. Havre}, \bibinfo{person}{B. Hetzler},
  {and} \bibinfo{person}{L. Nowell}.} \bibinfo{year}{2000}\natexlab{}.
\newblock \showarticletitle{ThemeRiver: visualizing theme changes over time}.
  In \bibinfo{booktitle}{\emph{IEEE Symposium on Information Visualization
  2000. INFOVIS 2000. Proceedings}}. \bibinfo{pages}{115--123}.
\newblock
\href{https://doi.org/10.1109/INFVIS.2000.885098}{doi:\nolinkurl{10.1109/INFVIS.2000.885098}}


\bibitem[Hernandez-Bocanegra and Ziegler(2023)]%
        {Hernandez-Bocanegra:2023:ExplainingRecsInConvo}
\bibfield{author}{\bibinfo{person}{Diana~C. Hernandez-Bocanegra} {and}
  \bibinfo{person}{J\"{u}rgen Ziegler}.} \bibinfo{year}{2023}\natexlab{}.
\newblock \showarticletitle{Explaining Recommendations through Conversations:
  Dialog Model and the Effects of Interface Type and Degree of Interactivity}.
\newblock \bibinfo{journal}{\emph{ACM Trans. Interact. Intell. Syst.}}
  \bibinfo{volume}{13}, \bibinfo{number}{2}, Article \bibinfo{articleno}{6}
  (\bibinfo{date}{April} \bibinfo{year}{2023}), \bibinfo{numpages}{47}~pages.
\newblock
\showISSN{2160-6455}
\href{https://doi.org/10.1145/3579541}{doi:\nolinkurl{10.1145/3579541}}


\bibitem[Higashinaka et~al\mbox{.}(2021)]%
        {Higashinaka:2021:TaxonomyErrorsDialogue}
\bibfield{author}{\bibinfo{person}{Ryuichiro Higashinaka},
  \bibinfo{person}{Masahiro Araki}, \bibinfo{person}{Hiroshi Tsukahara}, {and}
  \bibinfo{person}{Masahiro Mizukami}.} \bibinfo{year}{2021}\natexlab{}.
\newblock \showarticletitle{Integrated taxonomy of errors in chat-oriented
  dialogue systems}. In \bibinfo{booktitle}{\emph{Proceedings of the 22nd
  Annual Meeting of the Special Interest Group on Discourse and Dialogue}},
  \bibfield{editor}{\bibinfo{person}{Haizhou Li}, \bibinfo{person}{Gina-Anne
  Levow}, \bibinfo{person}{Zhou Yu}, \bibinfo{person}{Chitralekha Gupta},
  \bibinfo{person}{Berrak Sisman}, \bibinfo{person}{Siqi Cai},
  \bibinfo{person}{David Vandyke}, \bibinfo{person}{Nina Dethlefs},
  \bibinfo{person}{Yan Wu}, {and} \bibinfo{person}{Junyi~Jessy Li}} (Eds.).
  \bibinfo{publisher}{Association for Computational Linguistics},
  \bibinfo{address}{Singapore and Online}, \bibinfo{pages}{89--98}.
\newblock
\href{https://doi.org/10.18653/v1/2021.sigdial-1.10}{doi:\nolinkurl{10.18653/v1/2021.sigdial-1.10}}


\bibitem[Hong and Crisan(2023)]%
        {Hong:2023:ConversationalAIThreads}
\bibfield{author}{\bibinfo{person}{Matt-Heun Hong} {and}
  \bibinfo{person}{Anamaria Crisan}.} \bibinfo{year}{2023}\natexlab{}.
\newblock \bibinfo{title}{Conversational AI Threads for Visualizing
  Multidimensional Datasets}.
\newblock
\showeprint[arxiv]{2311.05590}~[cs.HC]
\urldef\tempurl%
\url{https://arxiv.org/abs/2311.05590}
\showURL{%
\tempurl}


\bibitem[Huang et~al\mbox{.}(2024)]%
        {Huang:2024:ConvLengthLLMStudy}
\bibfield{author}{\bibinfo{person}{Shih-Hong Huang}, \bibinfo{person}{Ya-Fang
  Lin}, \bibinfo{person}{Zeyu He}, \bibinfo{person}{Chieh-Yang Huang}, {and}
  \bibinfo{person}{Ting-Hao~Kenneth Huang}.} \bibinfo{year}{2024}\natexlab{}.
\newblock \showarticletitle{How Does Conversation Length Impact User’s
  Satisfaction? A Case Study of Length-Controlled Conversations with
  LLM-Powered Chatbots}. In \bibinfo{booktitle}{\emph{Extended Abstracts of the
  2024 CHI Conference on Human Factors in Computing Systems}}
  \emph{(\bibinfo{series}{CHI EA '24})}. \bibinfo{publisher}{Association for
  Computing Machinery}, \bibinfo{address}{New York, NY, USA}, Article
  \bibinfo{articleno}{188}, \bibinfo{numpages}{13}~pages.
\newblock
\showISBNx{9798400703317}
\href{https://doi.org/10.1145/3613905.3650823}{doi:\nolinkurl{10.1145/3613905.3650823}}


\bibitem[Jiang et~al\mbox{.}(2023)]%
        {Jiang:2023:Graphologue}
\bibfield{author}{\bibinfo{person}{Peiling Jiang}, \bibinfo{person}{Jude
  Rayan}, \bibinfo{person}{Steven~P. Dow}, {and} \bibinfo{person}{Haijun Xia}.}
  \bibinfo{year}{2023}\natexlab{}.
\newblock \showarticletitle{Graphologue: Exploring Large Language Model
  Responses with Interactive Diagrams}. In
  \bibinfo{booktitle}{\emph{Proceedings of the 36th Annual ACM Symposium on
  User Interface Software and Technology}} (San Francisco, CA, USA)
  \emph{(\bibinfo{series}{UIST '23})}. \bibinfo{publisher}{Association for
  Computing Machinery}, \bibinfo{address}{New York, NY, USA}, Article
  \bibinfo{articleno}{3}, \bibinfo{numpages}{20}~pages.
\newblock
\showISBNx{9798400701320}
\href{https://doi.org/10.1145/3586183.3606737}{doi:\nolinkurl{10.1145/3586183.3606737}}


\bibitem[Khot et~al\mbox{.}(2022)]%
        {khot2022decomposed}
\bibfield{author}{\bibinfo{person}{Tushar Khot}, \bibinfo{person}{Harsh
  Trivedi}, \bibinfo{person}{Matthew Finlayson}, \bibinfo{person}{Yao Fu},
  \bibinfo{person}{Kyle Richardson}, \bibinfo{person}{Peter Clark}, {and}
  \bibinfo{person}{Ashish Sabharwal}.} \bibinfo{year}{2022}\natexlab{}.
\newblock \showarticletitle{Decomposed prompting: A modular approach for
  solving complex tasks}.
\newblock \bibinfo{journal}{\emph{arXiv preprint arXiv:2210.02406}}
  (\bibinfo{year}{2022}).
\newblock


\bibitem[Kim et~al\mbox{.}(2021)]%
        {Kim:2021:ReviewDyadicConvVis}
\bibfield{author}{\bibinfo{person}{Joshua~Y. Kim}, \bibinfo{person}{Rafael~A.
  Calvo}, \bibinfo{person}{N.~J. Enfield}, {and} \bibinfo{person}{Kalina
  Yacef}.} \bibinfo{year}{2021}\natexlab{}.
\newblock \showarticletitle{A Systematic Review on Dyadic Conversation
  Visualizations}. In \bibinfo{booktitle}{\emph{Companion Publication of the
  2021 International Conference on Multimodal Interaction}} (Montreal, QC,
  Canada) \emph{(\bibinfo{series}{ICMI '21 Companion})}.
  \bibinfo{publisher}{Association for Computing Machinery},
  \bibinfo{address}{New York, NY, USA}, \bibinfo{pages}{137–147}.
\newblock
\showISBNx{9781450384711}
\href{https://doi.org/10.1145/3461615.3485396}{doi:\nolinkurl{10.1145/3461615.3485396}}


\bibitem[Kim et~al\mbox{.}(2024)]%
        {Kim:2024:DissatisfactionGPTTypes}
\bibfield{author}{\bibinfo{person}{Yoonsu Kim}, \bibinfo{person}{Jueon Lee},
  \bibinfo{person}{Seoyoung Kim}, \bibinfo{person}{Jaehyuk Park}, {and}
  \bibinfo{person}{Juho Kim}.} \bibinfo{year}{2024}\natexlab{}.
\newblock \showarticletitle{Understanding Users’ Dissatisfaction with ChatGPT
  Responses: Types, Resolving Tactics, and the Effect of Knowledge Level}. In
  \bibinfo{booktitle}{\emph{Proceedings of the 29th International Conference on
  Intelligent User Interfaces}} (Greenville, SC, USA)
  \emph{(\bibinfo{series}{IUI '24})}. \bibinfo{publisher}{Association for
  Computing Machinery}, \bibinfo{address}{New York, NY, USA},
  \bibinfo{pages}{385–404}.
\newblock
\showISBNx{9798400705083}
\href{https://doi.org/10.1145/3640543.3645148}{doi:\nolinkurl{10.1145/3640543.3645148}}


\bibitem[Kulkarni et~al\mbox{.}(2024)]%
        {kulkarni2024synthdst}
\bibfield{author}{\bibinfo{person}{Atharva Kulkarni},
  \bibinfo{person}{Bo-Hsiang Tseng}, \bibinfo{person}{Joel Moniz},
  \bibinfo{person}{Dhivya Piraviperumal}, \bibinfo{person}{Hong Yu}, {and}
  \bibinfo{person}{Shruti Bhargava}.} \bibinfo{year}{2024}\natexlab{}.
\newblock \showarticletitle{SynthDST: Synthetic Data is All You Need for
  Few-Shot Dialog State Tracking}. In \bibinfo{booktitle}{\emph{Proceedings of
  the 18th Conference of the European Chapter of the Association for
  Computational Linguistics (Volume 1: Long Papers)}}.
  \bibinfo{pages}{1988--2001}.
\newblock


\bibitem[Laban et~al\mbox{.}(2024)]%
        {laban2024beyond}
\bibfield{author}{\bibinfo{person}{Philippe Laban}, \bibinfo{person}{Jesse
  Vig}, \bibinfo{person}{Marti Hearst}, \bibinfo{person}{Caiming Xiong}, {and}
  \bibinfo{person}{Chien-Sheng Wu}.} \bibinfo{year}{2024}\natexlab{}.
\newblock \showarticletitle{Beyond the chat: Executable and verifiable
  text-editing with llms}. In \bibinfo{booktitle}{\emph{Proceedings of the 37th
  Annual ACM Symposium on User Interface Software and Technology}}.
  \bibinfo{pages}{1--23}.
\newblock


\bibitem[Lee et~al\mbox{.}(2025)]%
        {Lee:2025:GenAICriticalThinking}
\bibfield{author}{\bibinfo{person}{Hao-Ping~(Hank) Lee},
  \bibinfo{person}{Advait Sarkar}, \bibinfo{person}{Lev Tankelevitch},
  \bibinfo{person}{Ian Drosos}, \bibinfo{person}{Sean Rintel},
  \bibinfo{person}{Richard Banks}, {and} \bibinfo{person}{Nicholas Wilson}.}
  \bibinfo{year}{2025}\natexlab{}.
\newblock \showarticletitle{The Impact of Generative AI on Critical Thinking:
  Self-Reported Reductions in Cognitive Effort and Confidence Effects From a
  Survey of Knowledge Workers}. In \bibinfo{booktitle}{\emph{Proceedings of the
  ACM CHI Conference on Human Factors in Computing Systems}}.
  \bibinfo{publisher}{ACM}.
\newblock
\href{https://doi.org/10.1145/3706598.3713778}{doi:\nolinkurl{10.1145/3706598.3713778}}


\bibitem[Li et~al\mbox{.}(2024)]%
        {li2024chathf}
\bibfield{author}{\bibinfo{person}{Andrew Li}, \bibinfo{person}{Zhenduo Wang},
  \bibinfo{person}{Ethan Mendes}, \bibinfo{person}{Duong~Minh Le},
  \bibinfo{person}{Wei Xu}, {and} \bibinfo{person}{Alan Ritter}.}
  \bibinfo{year}{2024}\natexlab{}.
\newblock \showarticletitle{ChatHF: Collecting Rich Human Feedback from
  Real-time Conversations}. In \bibinfo{booktitle}{\emph{Proceedings of the
  2024 Conference on Empirical Methods in Natural Language Processing: System
  Demonstrations}}. \bibinfo{pages}{270--279}.
\newblock


\bibitem[Liang et~al\mbox{.}(2023)]%
        {Liang:2023:UsabilityAIAssistants}
\bibfield{author}{\bibinfo{person}{Jenny~T Liang}, \bibinfo{person}{Chenyang
  Yang}, {and} \bibinfo{person}{Brad~A Myers}.}
  \bibinfo{year}{2023}\natexlab{}.
\newblock \showarticletitle{A Large-Scale Survey on the Usability of AI
  Programming Assistants: Successes and Challenges}.
\newblock \bibinfo{journal}{\emph{arXiv preprint arXiv:2303.17125}}
  (\bibinfo{year}{2023}).
\newblock


\bibitem[Liu et~al\mbox{.}(2024)]%
        {Liu:2024:LostInTheMiddle}
\bibfield{author}{\bibinfo{person}{Nelson~F. Liu}, \bibinfo{person}{Kevin Lin},
  \bibinfo{person}{John Hewitt}, \bibinfo{person}{Ashwin Paranjape},
  \bibinfo{person}{Michele Bevilacqua}, \bibinfo{person}{Fabio Petroni}, {and}
  \bibinfo{person}{Percy Liang}.} \bibinfo{year}{2024}\natexlab{}.
\newblock \showarticletitle{Lost in the Middle: How Language Models Use Long
  Contexts}.
\newblock \bibinfo{journal}{\emph{Transactions of the Association for
  Computational Linguistics}}  \bibinfo{volume}{12} (\bibinfo{year}{2024}),
  \bibinfo{pages}{157--173}.
\newblock
\href{https://doi.org/10.1162/tacl_a_00638}{doi:\nolinkurl{10.1162/tacl_a_00638}}


\bibitem[Mahmood et~al\mbox{.}(2023)]%
        {mahmood2023llm}
\bibfield{author}{\bibinfo{person}{Amama Mahmood}, \bibinfo{person}{Junxiang
  Wang}, \bibinfo{person}{Bingsheng Yao}, \bibinfo{person}{Dakuo Wang}, {and}
  \bibinfo{person}{Chien-Ming Huang}.} \bibinfo{year}{2023}\natexlab{}.
\newblock \showarticletitle{Llm-powered conversational voice assistants:
  Interaction patterns, opportunities, challenges, and design guidelines}.
\newblock \bibinfo{journal}{\emph{arXiv preprint arXiv:2309.13879}}
  (\bibinfo{year}{2023}).
\newblock


\bibitem[Masson et~al\mbox{.}(2024)]%
        {masson2024directgpt}
\bibfield{author}{\bibinfo{person}{Damien Masson}, \bibinfo{person}{Sylvain
  Malacria}, \bibinfo{person}{G{\'e}ry Casiez}, {and} \bibinfo{person}{Daniel
  Vogel}.} \bibinfo{year}{2024}\natexlab{}.
\newblock \showarticletitle{Directgpt: A direct manipulation interface to
  interact with large language models}. In
  \bibinfo{booktitle}{\emph{Proceedings of the 2024 CHI Conference on Human
  Factors in Computing Systems}}. \bibinfo{pages}{1--16}.
\newblock


\bibitem[McAleese et~al\mbox{.}(2024)]%
        {McAleese:2024:CriticGPT}
\bibfield{author}{\bibinfo{person}{Nat McAleese}, \bibinfo{person}{Rai~Michael
  Pokorny}, \bibinfo{person}{Juan Felipe~Ceron Uribe}, \bibinfo{person}{Evgenia
  Nitishinskaya}, \bibinfo{person}{Maja Trebacz}, {and} \bibinfo{person}{Jan
  Leike}.} \bibinfo{year}{2024}\natexlab{}.
\newblock \showarticletitle{LLM Critics Help Catch LLM Bugs}.
\newblock \bibinfo{journal}{\emph{arXiv preprint arXiv:2407.00215}}
  (\bibinfo{year}{2024}).
\newblock


\bibitem[Mishra et~al\mbox{.}(2023)]%
        {Mishra:2023:PromptAid}
\bibfield{author}{\bibinfo{person}{Aditi Mishra}, \bibinfo{person}{Utkarsh
  Soni}, \bibinfo{person}{Anjana Arunkumar}, \bibinfo{person}{Jinbin Huang},
  \bibinfo{person}{Bum~Chul Kwon}, {and} \bibinfo{person}{Chris Bryan}.}
  \bibinfo{year}{2023}\natexlab{}.
\newblock \showarticletitle{Promptaid: Prompt exploration, perturbation,
  testing and iteration using visual analytics for large language models}.
\newblock \bibinfo{journal}{\emph{arXiv preprint arXiv:2304.01964}}
  (\bibinfo{year}{2023}).
\newblock


\bibitem[Niu et~al\mbox{.}(2024)]%
        {niu2024enhancing}
\bibfield{author}{\bibinfo{person}{Cheng Niu}, \bibinfo{person}{Xingguang
  Wang}, \bibinfo{person}{Xuxin Cheng}, \bibinfo{person}{Juntong Song}, {and}
  \bibinfo{person}{Tong Zhang}.} \bibinfo{year}{2024}\natexlab{}.
\newblock \showarticletitle{Enhancing Dialogue State Tracking Models through
  LLM-backed User-Agents Simulation}. In \bibinfo{booktitle}{\emph{Proceedings
  of the 62nd Annual Meeting of the Association for Computational Linguistics
  (Volume 1: Long Papers)}}. \bibinfo{pages}{8724--8741}.
\newblock


\bibitem[Papenmeier et~al\mbox{.}(2022)]%
        {Papenmeier:2022:TrustAIExplanations}
\bibfield{author}{\bibinfo{person}{Andrea Papenmeier}, \bibinfo{person}{Dagmar
  Kern}, \bibinfo{person}{Gwenn Englebienne}, {and} \bibinfo{person}{Christin
  Seifert}.} \bibinfo{year}{2022}\natexlab{}.
\newblock \showarticletitle{It’s Complicated: The Relationship between User
  Trust, Model Accuracy and Explanations in AI}.
\newblock \bibinfo{journal}{\emph{ACM Trans. Comput.-Hum. Interact.}}
  \bibinfo{volume}{29}, \bibinfo{number}{4}, Article \bibinfo{articleno}{35}
  (\bibinfo{date}{March} \bibinfo{year}{2022}), \bibinfo{numpages}{33}~pages.
\newblock
\showISSN{1073-0516}
\href{https://doi.org/10.1145/3495013}{doi:\nolinkurl{10.1145/3495013}}


\bibitem[Ross et~al\mbox{.}(2023)]%
        {Ross:2023:ProgrammersAssistant}
\bibfield{author}{\bibinfo{person}{Steven~I. Ross}, \bibinfo{person}{Fernando
  Martinez}, \bibinfo{person}{Stephanie Houde}, \bibinfo{person}{Michael
  Muller}, {and} \bibinfo{person}{Justin~D. Weisz}.}
  \bibinfo{year}{2023}\natexlab{}.
\newblock \showarticletitle{The Programmer’s Assistant: Conversational
  Interaction with a Large Language Model for Software Development}. In
  \bibinfo{booktitle}{\emph{Proceedings of the 28th International Conference on
  Intelligent User Interfaces}} (Sydney, NSW, Australia)
  \emph{(\bibinfo{series}{IUI '23})}. \bibinfo{publisher}{Association for
  Computing Machinery}, \bibinfo{address}{New York, NY, USA},
  \bibinfo{pages}{491–514}.
\newblock
\showISBNx{9798400701061}
\href{https://doi.org/10.1145/3581641.3584037}{doi:\nolinkurl{10.1145/3581641.3584037}}


\bibitem[Shi et~al\mbox{.}(2023)]%
        {Shi:2023:LLMsDistracted}
\bibfield{author}{\bibinfo{person}{Freda Shi}, \bibinfo{person}{Xinyun Chen},
  \bibinfo{person}{Kanishka Misra}, \bibinfo{person}{Nathan Scales},
  \bibinfo{person}{David Dohan}, \bibinfo{person}{Ed~H. Chi},
  \bibinfo{person}{Nathanael Sch\"{a}rli}, {and} \bibinfo{person}{Denny Zhou}.}
  \bibinfo{year}{2023}\natexlab{}.
\newblock \showarticletitle{Large Language Models Can Be Easily Distracted by
  Irrelevant Context}. In \bibinfo{booktitle}{\emph{Proceedings of the 40th
  International Conference on Machine Learning}}
  \emph{(\bibinfo{series}{Proceedings of Machine Learning Research},
  Vol.~\bibinfo{volume}{202})}, \bibfield{editor}{\bibinfo{person}{Andreas
  Krause}, \bibinfo{person}{Emma Brunskill}, \bibinfo{person}{Kyunghyun Cho},
  \bibinfo{person}{Barbara Engelhardt}, \bibinfo{person}{Sivan Sabato}, {and}
  \bibinfo{person}{Jonathan Scarlett}} (Eds.). \bibinfo{publisher}{PMLR},
  \bibinfo{pages}{31210--31227}.
\newblock
\urldef\tempurl%
\url{https://proceedings.mlr.press/v202/shi23a.html}
\showURL{%
\tempurl}


\bibitem[Shi et~al\mbox{.}(2024)]%
        {shi2024wildfeedback}
\bibfield{author}{\bibinfo{person}{Taiwei Shi}, \bibinfo{person}{Zhuoer Wang},
  \bibinfo{person}{Longqi Yang}, \bibinfo{person}{Ying-Chun Lin},
  \bibinfo{person}{Zexue He}, \bibinfo{person}{Mengting Wan},
  \bibinfo{person}{Pei Zhou}, \bibinfo{person}{Sujay Jauhar},
  \bibinfo{person}{Sihao Chen}, \bibinfo{person}{Shan Xia}, {et~al\mbox{.}}}
  \bibinfo{year}{2024}\natexlab{}.
\newblock \showarticletitle{Wildfeedback: Aligning llms with in-situ user
  interactions and feedback}.
\newblock \bibinfo{journal}{\emph{arXiv preprint arXiv:2408.15549}}
  (\bibinfo{year}{2024}).
\newblock


\bibitem[Stigall et~al\mbox{.}(2023)]%
        {Stigall:2023:ChatbotsVisualEDA}
\bibfield{author}{\bibinfo{person}{Brodrick Stigall}, \bibinfo{person}{Ryan
  Rossi}, \bibinfo{person}{Jane Hoffswell}, \bibinfo{person}{Xiang Chen},
  \bibinfo{person}{Shunan Guo}, \bibinfo{person}{Fan Du},
  \bibinfo{person}{Eunyee Koh}, {and} \bibinfo{person}{Kelly Caine}.}
  \bibinfo{year}{2023}\natexlab{}.
\newblock \showarticletitle{On Chatbots for Visual Exploratory Data Analysis}.
  In \bibinfo{booktitle}{\emph{2023 IEEE International Conference on Big Data
  (BigData)}}. \bibinfo{pages}{5924--5929}.
\newblock
\href{https://doi.org/10.1109/BigData59044.2023.10386335}{doi:\nolinkurl{10.1109/BigData59044.2023.10386335}}


\bibitem[Stivers et~al\mbox{.}(2009)]%
        {Stivers:2009:TurnTaking}
\bibfield{author}{\bibinfo{person}{Tanya Stivers}, \bibinfo{person}{N~J
  Enfield}, \bibinfo{person}{Penelope Brown}, \bibinfo{person}{Christina
  Englert}, \bibinfo{person}{Makoto Hayashi}, \bibinfo{person}{Trine
  Heinemann}, \bibinfo{person}{Gertie Hoymann}, \bibinfo{person}{Federico
  Rossano}, \bibinfo{person}{Jan~Peter de Ruiter}, \bibinfo{person}{Kyung-Eun
  Yoon}, {and} \bibinfo{person}{Stephen~C Levinson}.}
  \bibinfo{year}{2009}\natexlab{}.
\newblock \showarticletitle{Universals and cultural variation in turn-taking in
  conversation}.
\newblock \bibinfo{journal}{\emph{Proc. Natl. Acad. Sci. U. S. A.}}
  \bibinfo{volume}{106}, \bibinfo{number}{26} (\bibinfo{date}{June}
  \bibinfo{year}{2009}), \bibinfo{pages}{10587--10592}.
\newblock


\bibitem[Subramonyam et~al\mbox{.}(2024)]%
        {subramonyam2023bridging}
\bibfield{author}{\bibinfo{person}{Hari Subramonyam}, \bibinfo{person}{Roy
  Pea}, \bibinfo{person}{Christopher Pondoc}, \bibinfo{person}{Maneesh
  Agrawala}, {and} \bibinfo{person}{Colleen Seifert}.}
  \bibinfo{year}{2024}\natexlab{}.
\newblock \showarticletitle{Bridging the Gulf of Envisioning: Cognitive
  Challenges in Prompt Based Interactions with LLMs}. In
  \bibinfo{booktitle}{\emph{Proceedings of the 2024 CHI Conference on Human
  Factors in Computing Systems}} (Honolulu, HI, USA)
  \emph{(\bibinfo{series}{CHI '24})}. \bibinfo{publisher}{Association for
  Computing Machinery}, \bibinfo{address}{New York, NY, USA}, Article
  \bibinfo{articleno}{1039}, \bibinfo{numpages}{19}~pages.
\newblock
\showISBNx{9798400703300}
\href{https://doi.org/10.1145/3613904.3642754}{doi:\nolinkurl{10.1145/3613904.3642754}}


\bibitem[Suchmann et~al\mbox{.}(2023)]%
        {Suchmann:2023:BranchingConvoInSituVis}
\bibfield{author}{\bibinfo{person}{Lovis~Bero Suchmann},
  \bibinfo{person}{Nicole Kr\"{a}mer}, {and} \bibinfo{person}{J\"{u}rgen
  Ziegler}.} \bibinfo{year}{2023}\natexlab{}.
\newblock \showarticletitle{Branching Preferences: Visualizing Non-linear Topic
  Progression in Conversational Recommender Systems}. In
  \bibinfo{booktitle}{\emph{Adjunct Proceedings of the 31st ACM Conference on
  User Modeling, Adaptation and Personalization}} (Limassol, Cyprus)
  \emph{(\bibinfo{series}{UMAP '23 Adjunct})}. \bibinfo{publisher}{Association
  for Computing Machinery}, \bibinfo{address}{New York, NY, USA},
  \bibinfo{pages}{199–205}.
\newblock
\showISBNx{9781450398916}
\href{https://doi.org/10.1145/3563359.3597380}{doi:\nolinkurl{10.1145/3563359.3597380}}


\bibitem[Suh et~al\mbox{.}(2023)]%
        {Suh:2023:Sensecape}
\bibfield{author}{\bibinfo{person}{Sangho Suh}, \bibinfo{person}{Bryan Min},
  \bibinfo{person}{Srishti Palani}, {and} \bibinfo{person}{Haijun Xia}.}
  \bibinfo{year}{2023}\natexlab{}.
\newblock \showarticletitle{Sensecape: Enabling Multilevel Exploration and
  Sensemaking with Large Language Models}. In
  \bibinfo{booktitle}{\emph{Proceedings of the 36th Annual ACM Symposium on
  User Interface Software and Technology}} (San Francisco, CA, USA)
  \emph{(\bibinfo{series}{UIST '23})}. \bibinfo{publisher}{Association for
  Computing Machinery}, \bibinfo{address}{New York, NY, USA}, Article
  \bibinfo{articleno}{1}, \bibinfo{numpages}{18}~pages.
\newblock
\showISBNx{9798400701320}
\href{https://doi.org/10.1145/3586183.3606756}{doi:\nolinkurl{10.1145/3586183.3606756}}


\bibitem[Tankelevitch et~al\mbox{.}(2024)]%
        {tankelevitch2024metacognitive}
\bibfield{author}{\bibinfo{person}{Lev Tankelevitch}, \bibinfo{person}{Viktor
  Kewenig}, \bibinfo{person}{Auste Simkute}, \bibinfo{person}{Ava~Elizabeth
  Scott}, \bibinfo{person}{Advait Sarkar}, \bibinfo{person}{Abigail Sellen},
  {and} \bibinfo{person}{Sean Rintel}.} \bibinfo{year}{2024}\natexlab{}.
\newblock \showarticletitle{The metacognitive demands and opportunities of
  generative AI}. In \bibinfo{booktitle}{\emph{Proceedings of the 2024 CHI
  Conference on Human Factors in Computing Systems}}. \bibinfo{pages}{1--24}.
\newblock


\bibitem[Tat and Carpendale(2002)]%
        {Tat:2002:VisualizingHumanDialogue}
\bibfield{author}{\bibinfo{person}{A. Tat} {and} \bibinfo{person}{M.S.T.
  Carpendale}.} \bibinfo{year}{2002}\natexlab{}.
\newblock \showarticletitle{Visualising human dialog}. In
  \bibinfo{booktitle}{\emph{Proceedings Sixth International Conference on
  Information Visualisation}}. \bibinfo{pages}{16--21}.
\newblock
\href{https://doi.org/10.1109/IV.2002.1028751}{doi:\nolinkurl{10.1109/IV.2002.1028751}}


\bibitem[Tufte and Graves-Morris(1983)]%
        {Tufte:1983:VisualDisplayInfo}
\bibfield{author}{\bibinfo{person}{Edward~R Tufte} {and}
  \bibinfo{person}{Peter~R Graves-Morris}.} \bibinfo{year}{1983}\natexlab{}.
\newblock \bibinfo{booktitle}{\emph{The visual display of quantitative
  information}}. Vol.~\bibinfo{volume}{2}.
\newblock \bibinfo{publisher}{Graphics press Cheshire, CT}.
\newblock


\bibitem[Venolia and Neustaedter(2003)]%
        {Venolia:2003:EmailConversationVis}
\bibfield{author}{\bibinfo{person}{Gina~Danielle Venolia} {and}
  \bibinfo{person}{Carman Neustaedter}.} \bibinfo{year}{2003}\natexlab{}.
\newblock \showarticletitle{Understanding sequence and reply relationships
  within email conversations: a mixed-model visualization}. In
  \bibinfo{booktitle}{\emph{Proceedings of the SIGCHI Conference on Human
  Factors in Computing Systems}} (Ft. Lauderdale, Florida, USA)
  \emph{(\bibinfo{series}{CHI '03})}. \bibinfo{publisher}{Association for
  Computing Machinery}, \bibinfo{address}{New York, NY, USA},
  \bibinfo{pages}{361–368}.
\newblock
\showISBNx{1581136307}
\href{https://doi.org/10.1145/642611.642674}{doi:\nolinkurl{10.1145/642611.642674}}


\bibitem[Wang et~al\mbox{.}(2024)]%
        {wang2024task}
\bibfield{author}{\bibinfo{person}{Ben Wang}, \bibinfo{person}{Jiqun Liu},
  \bibinfo{person}{Jamshed Karimnazarov}, {and} \bibinfo{person}{Nicolas
  Thompson}.} \bibinfo{year}{2024}\natexlab{}.
\newblock \showarticletitle{Task supportive and personalized human-large
  language model interaction: A user study}. In
  \bibinfo{booktitle}{\emph{Proceedings of the 2024 Conference on Human
  Information Interaction and Retrieval}}. \bibinfo{pages}{370--375}.
\newblock


\bibitem[Wang et~al\mbox{.}(2021)]%
        {Wang:2021:DiscussionFlows}
\bibfield{author}{\bibinfo{person}{Tao Wang}, \bibinfo{person}{Mandy Keck},
  {and} \bibinfo{person}{Zana Vosough}.} \bibinfo{year}{2021}\natexlab{}.
\newblock \showarticletitle{{Discussion Flows: An Interactive Visualization for
  Analyzing Engagement in Multi-Party Meetings}}. In
  \bibinfo{booktitle}{\emph{EuroVis 2021 - Short Papers}},
  \bibfield{editor}{\bibinfo{person}{Marco Agus}, \bibinfo{person}{Christoph
  Garth}, {and} \bibinfo{person}{Andreas Kerren}} (Eds.).
  \bibinfo{publisher}{The Eurographics Association}.
\newblock
\showISBNx{978-3-03868-143-4}
\href{https://doi.org/10.2312/evs.20211060}{doi:\nolinkurl{10.2312/evs.20211060}}


\bibitem[Wei et~al\mbox{.}(2022)]%
        {wei2022chain}
\bibfield{author}{\bibinfo{person}{Jason Wei}, \bibinfo{person}{Xuezhi Wang},
  \bibinfo{person}{Dale Schuurmans}, \bibinfo{person}{Maarten Bosma},
  \bibinfo{person}{Fei Xia}, \bibinfo{person}{Ed Chi}, \bibinfo{person}{Quoc~V
  Le}, \bibinfo{person}{Denny Zhou}, {et~al\mbox{.}}}
  \bibinfo{year}{2022}\natexlab{}.
\newblock \showarticletitle{Chain-of-thought prompting elicits reasoning in
  large language models}.
\newblock \bibinfo{journal}{\emph{Advances in neural information processing
  systems}}  \bibinfo{volume}{35} (\bibinfo{year}{2022}),
  \bibinfo{pages}{24824--24837}.
\newblock


\bibitem[Wu et~al\mbox{.}(2023)]%
        {Wu:2023:ScatterShot}
\bibfield{author}{\bibinfo{person}{Sherry Wu}, \bibinfo{person}{Hua Shen},
  \bibinfo{person}{Daniel~S Weld}, \bibinfo{person}{Jeffrey Heer}, {and}
  \bibinfo{person}{Marco~Tulio Ribeiro}.} \bibinfo{year}{2023}\natexlab{}.
\newblock \showarticletitle{ScatterShot: Interactive In-context Example
  Curation for Text Transformation}. In \bibinfo{booktitle}{\emph{Proceedings
  of the 28th International Conference on Intelligent User Interfaces}}
  (Sydney, NSW, Australia) \emph{(\bibinfo{series}{IUI '23})}.
  \bibinfo{publisher}{Association for Computing Machinery},
  \bibinfo{address}{New York, NY, USA}, \bibinfo{pages}{353–367}.
\newblock
\showISBNx{9798400701061}
\href{https://doi.org/10.1145/3581641.3584059}{doi:\nolinkurl{10.1145/3581641.3584059}}


\bibitem[Wu et~al\mbox{.}(2022)]%
        {Tongshuang:2022:PromptChainer}
\bibfield{author}{\bibinfo{person}{Tongshuang Wu}, \bibinfo{person}{Ellen
  Jiang}, \bibinfo{person}{Aaron Donsbach}, \bibinfo{person}{Jeff Gray},
  \bibinfo{person}{Alejandra Molina}, \bibinfo{person}{Michael Terry}, {and}
  \bibinfo{person}{Carrie~J Cai}.} \bibinfo{year}{2022}\natexlab{}.
\newblock \showarticletitle{PromptChainer: Chaining Large Language Model
  Prompts through Visual Programming}. In \bibinfo{booktitle}{\emph{Extended
  Abstracts of the 2022 CHI Conference on Human Factors in Computing Systems}}
  (New Orleans, LA, USA) \emph{(\bibinfo{series}{CHI EA '22})}.
  \bibinfo{publisher}{Association for Computing Machinery},
  \bibinfo{address}{New York, NY, USA}, Article \bibinfo{articleno}{359},
  \bibinfo{numpages}{10}~pages.
\newblock
\showISBNx{9781450391566}
\href{https://doi.org/10.1145/3491101.3519729}{doi:\nolinkurl{10.1145/3491101.3519729}}


\bibitem[Zamfirescu-Pereira et~al\mbox{.}(2023)]%
        {zamfirescu2023johnny}
\bibfield{author}{\bibinfo{person}{J~Diego Zamfirescu-Pereira},
  \bibinfo{person}{Richmond~Y Wong}, \bibinfo{person}{Bjoern Hartmann}, {and}
  \bibinfo{person}{Qian Yang}.} \bibinfo{year}{2023}\natexlab{}.
\newblock \showarticletitle{Why Johnny can’t prompt: how non-AI experts try
  (and fail) to design LLM prompts}. In \bibinfo{booktitle}{\emph{Proceedings
  of the 2023 CHI conference on human factors in computing systems}}.
  \bibinfo{pages}{1--21}.
\newblock


\bibitem[Zheng et~al\mbox{.}(2023)]%
        {Zheng:2023:ChatbotArena}
\bibfield{author}{\bibinfo{person}{Lianmin Zheng}, \bibinfo{person}{Wei-Lin
  Chiang}, \bibinfo{person}{Ying Sheng}, \bibinfo{person}{Siyuan Zhuang},
  \bibinfo{person}{Zhanghao Wu}, \bibinfo{person}{Yonghao Zhuang},
  \bibinfo{person}{Zi Lin}, \bibinfo{person}{Zhuohan Li},
  \bibinfo{person}{Dacheng Li}, \bibinfo{person}{Eric Xing},
  \bibinfo{person}{Hao Zhang}, \bibinfo{person}{Joseph~E Gonzalez}, {and}
  \bibinfo{person}{Ion Stoica}.} \bibinfo{year}{2023}\natexlab{}.
\newblock \showarticletitle{Judging LLM-as-a-Judge with MT-Bench and Chatbot
  Arena}. In \bibinfo{booktitle}{\emph{Advances in Neural Information
  Processing Systems}}, \bibfield{editor}{\bibinfo{person}{A.~Oh},
  \bibinfo{person}{T.~Naumann}, \bibinfo{person}{A.~Globerson},
  \bibinfo{person}{K.~Saenko}, \bibinfo{person}{M.~Hardt}, {and}
  \bibinfo{person}{S.~Levine}} (Eds.), Vol.~\bibinfo{volume}{36}.
  \bibinfo{publisher}{Curran Associates, Inc.}, \bibinfo{pages}{46595--46623}.
\newblock
\urldef\tempurl%
\url{https://proceedings.neurips.cc/paper_files/paper/2023/file/91f18a1287b398d378ef22505bf41832-Paper-Datasets_and_Benchmarks.pdf}
\showURL{%
\tempurl}


\end{thebibliography}

\appendix

\section{LLM Prompts}
\label{sec:llm_prompts}

\subsection{Infer goals}
\label{sec:prompt_infer}

\begin{lstlisting}[breaklines]
You will be presented with human dialogue in a conversation with you, an assistant. Your task is to extract every clause verbatim from the document exactly as it appears.

List all clauses in the dialogue that are either a question, request, offer, or suggestion. Briefly summarize how to address the goal of the clause in ONE sentence.

Please respond ONLY with a valid JSON in the following format:

{{
    "clauses": [ 
        {{"clause": "<CLAUSE_1>", "type": "<TYPE_1>", "summary": "<SUMMARY_1>"}},
        {{"clause": "<CLAUSE_2>", "type": "<TYPE_2>", "summary": "<SUMMARY_2>"}},
    ]
}}
\end{lstlisting}

\subsection{Merge goals}
\label{sec:prompt_merge}

Input parameters:

\begin{itemize}
    \item \verb|old_goals_str_list| --- A newline-delimited string list of numbered goals. Example: ``1. ... \textbackslash{}n 2. ...''
    \item \verb|new_goals_str_list| --- A newline-delimited string list of numbered goals. Example: ``1. ... \textbackslash{}n 2. ...''
\end{itemize}

\begin{lstlisting}[breaklines]
You have one set of old numbered bullet point goals:
{old_goals_str_list}

You have another set of new numbered bullet point goals:
{new_goals_str_list}

Merge the two lists of bullet point goals into a single updated list of goals. Use the following three operations as rules to perform the merge:

* Replace: If a new goal contradicts an old goal, replace the old goal with the new goal. List the number of the old goal, then the number of the new goal.
* Combine: If a new goal is similar to an old goal, combine the old goal and the new goal into a new combined goal. List the number of the old goal, then the number of the new goal.
* Keep: If a goal is unique, keep that goal in the updated list. List the original number of the goal.

Please respond ONLY with a valid JSON in the following format:

{{
    "operations": [
        {{
            "updated_goal": "<GOAL_1>",
            "operation": "<OPERATION_1>", 
            "goal_numbers": ["<GOAL_NUMBER_1>", "<GOAL_NUMBER_2>"]
        }},
        {{
            "updated_goal": "<GOAL_2>",
            "operation": "<OPERATION_2>",
            "goal_numbers": ["<GOAL_NUMBER_1>", "<GOAL_NUMBER_2>"]
        }},
    ]
}}
\end{lstlisting}

\subsection{Evaluate goals}
\label{sec:prompt_evaluate}

Input parameters:

\begin{itemize}
    \item \verb|goal_str| --- A string representing a conversational goal. Example: ``Use figurative language.''
\end{itemize}

\begin{lstlisting}[breaklines]
You will be presented with human dialogue and a response from you, an assistant. Your task is to evaluate the assistant response in terms of the following conversational goal: {goal_str}

Categorize how the assistant response addresses the goal in one of three categories. The categories are confirm, contradict, or ignore. Explain the relationship between the response and the goal in ONE sentence. Extract clauses verbatim from the response exactly as they appear as examples that show evidence to support your explanation.

Please respond ONLY with a valid JSON in the following format:

{{
    "category": "<CATEGORY_1>",  
    "explanation": "<EXPLANATION_1>", 
    "examples": ["<EXAMPLE_1>", "<EXAMPLE_2>"]
}}
\end{lstlisting}

\subsection{Keyphrase extraction}
\label{sec:prompt_keyphrase}

\begin{lstlisting}[breaklines]
You will be given an assistant response. Your task is to extract every phrase verbatim from the response exactly as it appears.

List the phrases that capture the most salient topics of the response.

Please respond ONLY with a valid JSON in the following format:

{{
    "keyphrases": ["<KEYPHRASE_1>", "<KEYPHRASE_2>"]
}}
\end{lstlisting}

\end{document}